\documentclass[12pt,preprint]{aastex}

\begin{document}

\title{Bulge and Clump Evolution in Hubble Ultra Deep Field Clump
Clusters, Chains and Spiral Galaxies}

\author{Bruce G. Elmegreen}
\affil{IBM Research Division, T.J. Watson Research Center, P.O. Box
218, Yorktown Heights, NY 10598} \email{bge@watson.ibm.com}

\author{Debra Meloy Elmegreen}
\affil{Vassar College, Dept. of Physics \& Astronomy, Box 745, Poughkeepsie, NY 12604}
\email{elmegreen@vassar.edu}

\author{Maria Ximena Fernandez}
\affil{Vassar College, Dept. of Physics \& Astronomy, Box 745, Poughkeepsie, NY 12604}
\email{mafernandez@vassar.edu}

\author{Jenna Jo Lemonias}
\affil{Vassar College, Dept. of Physics \& Astronomy, Box 745, Poughkeepsie, NY 12604}
\email{jelemonias@vassar.edu}

\begin{abstract}
Clump clusters and chain galaxies in the Hubble Ultra Deep Field are
examined for bulges in the NICMOS images. Approximately 50\% of the
clump clusters and 30\% of the chains have relatively red and massive
clumps that could be young bulges. Magnitudes and colors are determined
for these bulge-like objects and for the bulges in spiral galaxies, and
for all of the prominent star-formation clumps in these three galaxy
types. The colors are fitted to population evolution models to
determine the bulge and clump masses, ages, star-formation rate decay
times, and extinctions.  The results indicate that bulge-like objects
in clump cluster and chain galaxies have similar ages and 2 to 5 times
larger masses compared to the star-formation clumps, while the bulges
in spirals have $\sim6$ times larger ages and 20 to 30 times larger
masses than the clumps. All systems appear to have an underlying red
disk population. The masses of star-forming clumps are typically in a
range from $10^7$ to $10^8\;M_\odot$; their ages have a wide range
around $\sim10^2$ Myr. Ages and extinctions both decrease with
redshift. Star formation is probably the result of gravitational
instabilities in the disk gas, in which case the large clump mass in
the UDF is the result of a high gas velocity dispersion, $30$ km
s$^{-1}$ or more, combined with a high gas mass column density,
$\sim100\;M_\odot$ pc$^{-2}$. Because clump clusters and chains
dominate disk galaxies beyond $z\sim1$, the observations suggest that
these types represent an early phase in the formation of modern spiral
galaxies, when the bulge and inner disk formed.
\end{abstract}


\section{Introduction}

Galaxies observed with the Advanced Camera for Surveys (ACS) on the
Hubble Space Telescope become increasingly clumpy and irregular beyond
redshift $z\sim1$, partly because star-forming regions get
intrinsically more massive at higher redshifts, partly because of
bandshifting of uv-bright regions into the optical bands, and partly
because of an increased importance of interactions in the young
universe (e.g., Abraham et al. 1996a,b; Conselice, Blackburne, \&
Papovich 2005; Elmegreen \& Elmegreen 2005; Lotz et al. 2006). In our
morphological catalog of galaxies (Elmegreen et al. 2005a) for the
Ultra Deep Field (UDF; Beckwith et al. 2006), clumpy structures are
present for all galaxy types, including ellipticals and spirals. The
most irregular systems are clump-clusters, which are oval collections
of bright clumps, and chain galaxies (Cowie, Hu, \& Songaila 1995),
which are linear alignments of bright clumps. The distribution of the
ratio of axes among clump cluster and chain galaxies suggests that the
chains are edge-on clump-clusters (Elmegreen,  Elmegreen \& Hirst 2004;
Elmegreen \& Elmegreen 2005). Some clump clusters have 40\% of their
$i_{775}$-band light in the clumps, although the average is 27\%
(Elmegreen et al. 2005). Clump masses measured from photometry range
from $10^7\;M_\odot$ to $10^9\;M_\odot$ in stars (e.g., Elmegreen \&
Elmegreen 2005; Elmegreen et al. 2007b); gas masses are not observed
yet. Some clump-cluster galaxies could be interacting systems (Overzier
et al. 2008), although the flattest chain galaxies seem to be
unperturbed disks (Elmegreen \& Elmegreen 2006).

Clump-clusters and chains that avoid destructive mergers presumably
evolve into modern disk galaxies. We would like to understand the time
sequence for their evolution and the processes that convert such
irregular systems into smooth disks with bulges and exponential
profiles. Numerical simulations of gas-rich, turbulent disks resembling
clump clusters suggest that clumps form by gravitational instabilities
and have the local Jeans mass (Bournaud, Elmegreen \& Elmegreen 2007),
which can exceed $10^8\;M_\odot$ at the high level of turbulence that
is observed (F{\"o}rster Schreiber et al. 2006; Genzel et al. 2006). If
the clumps are massive enough, they interact gravitationally, causing a
redistribution of angular momentum outward and an accretion of clump
and other disk mass inward. This process forms a bulge and exponential
disk within several orbit times, or $\sim1$ Gyr for a galaxy like the
Milky Way (Noguchi 1999; Immeli et al. 2004ab; Carollo et al. 2007;
Bournaud, Elmegreen, \& Elmegreen 2007; Bournaud et al. 2008;
Elmegreen, Bournaud \& Elmegreen 2008a,b).

Other rapid formation mechanisms for classical bulges include galaxy
mergers (Hernquist \& Barnes 1991), cluster mergers (Fu et al. 2003,
Zavala et al. 2008) and cosmological gas accretion (Steinmetz \& Muller
1995; Xu et al. 2007). Pseudobulges presumably formed more slowly by
vertical bar instabilities (Combes, \& Sanders 1981; van den Bosch
1998; Avila-Reese et al. 2005; Debattista et al. 2006; Athanassoula
2007, 2008), and torque-driven accretion (Pfenniger \& Norman 1990;
Zhang 1999; see reviews in Kormendy \& Kennicutt 2004; Conselice 2008;
Elmegreen, Bournaud, \& Elmegreen 2008b).

The purpose of this paper is to look for young bulges in clump cluster and
chain galaxies, and to try to understand how these galaxies fit into an
evolutionary sequence of disk systems. To do this, we measure and model
the magnitudes of all the prominent clumps in the clump-cluster, chain,
and spiral galaxies in the UDF. The models then give the clump ages,
the star formation decay times in each clump, the extinctions, and the
masses.

Section \ref{sect:data} describes the data and photometry, section
\ref{sect:morph} considers the morphologies of bulges in clump cluster
and chain galaxies, section \ref{sect:color} discusses the colors and
magnitudes of the galaxies, bulges and clumps, section
\ref{sect:models} presents models for intrinsic properties of the
bulges and clumps, and section \ref{sect:disc} discusses the results. A
conclusion is in section \ref{sect:conc}.

\section{Data on UDF Bulges and Clumps}
\label{sect:data}

The definition of a bulge is subjective for highly irregular galaxies
like chains and clump clusters. This contrasts to the case for spiral
galaxies at high redshift, which have clear bulges that are centralized
and relatively red. Most of the irregular galaxies considered here
either have nothing that could be identified as a bulge, even in Near
Infrared Camera Multi-Object Spectrometer (NICMOS) images where bulges
should be seen, or they have one clump that is slightly redder,
brighter, or more centrally positioned than the others and could
conceivably be a young bulge. We cannot place a quantitative measure on
what is considered to be a bulge because then our mass and age results
would reflect this initial selection.  Instead we view the images and
select the clump that is reddest, however red that is, or brightest,
and then examine the age and mass differences compared to the other
clumps. NICMOS is essential for this because by definition, chain and
clump cluster galaxies have no centralized bulges in optical images --
if they did, they would be classified as spirals. If there are no
particularly red clumps, or no particularly bright centralized clumps,
i.e., if all of the clumps look about the same in both optical and
NICMOS bands, and none are particularly concentrated toward the center,
then we conclude that there is no ``bulge-like clump'' (BLC) in that
galaxy. In fact, we find that either clump cluster and chain galaxies
have no BLCs, or, if they do, then those BLCs are generally more
similar to the other clumps in the same galaxies than are the bulges in
spiral galaxies compared to their clumps. This comparison is made on a
galaxy-by-galaxy basis to minimize uncertainties in redshift and
metallicity, which should be more uniform within a galaxy than between
galaxies.  We conclude below that the bulges in clump clusters and
chains are morphologically younger than the bulges in spirals. This
agrees with our direct age measurements in many cases, but because
galaxy formation may be spread out somewhat over redshift, and because
some spiral bulges may form late while some clump cluster galaxies may
form relatively early, there is no obvious one-to-one correspondence
between bulge age and bulge morphology at the redshifts of interest
here. With our definition of BLCs, there is at most one BLC per chain
or clump cluster galaxy, and this is the most massive of the red clumps
seen in the NICMOS images.

We use ACS images for the whole UDF field and NICMOS images in the
portion of the UDF covered by that instrument. The NICMOS images are
three times poorer resolution than the ACS images, but NICMOS allows us
to assess whether a galaxy contains an evolved bulge, which should be
red in restframe colors and prominent in NICMOS. While all of the
spiral galaxies contain an obvious central bulge in the ACS image (this
defines them as spiral types), none of the clump-clusters and chain
galaxies do. Still, we find BLCs in NICMOS for one-third of the
well-resolved chains and one-half of the well-resolved clump clusters.
In most cases, these BLCs are also seen as clumps in the ACS images,
although the overall structures of these galaxies are irregular (e.g.,
about 20\% of the BLCs in each irregular type are significantly offset
from the center), and the BLCs are bluer in clump clusters and chains
(compared to the rest of the galaxy) than are the bulges in spirals. We
also find prominent NICMOS emission from inner red disks that are not
likely to be bulges in the usual sense; we call this diffuse red
emission. We also use the Spitzer IRAC images that are available for a
portion of the UDF field.

The AB magnitudes were determined for all of the clumps, BLCs, and
bulges studied here using the four ACS passbands, $B_{435}$, $V_{606}$,
$i_{775}$, and $z_{850}$, and the two NICMOS passbands, $J_{1100}$ and
$H_{1600}$. Magnitudes were determined for all clumps, BLCs, and bulges
by defining boxes around the objects using the IRAF task $\it imstat$.
Their clump boundaries are defined at a level about 10$\sigma$ above
the noise. Zeropoint conversions for each filter are available through
the online handbooks. Magnitude measurement errors are estimated to be
about 0.1 mag. The $J$ and $H$ magnitudes of the bulges and diffuse
inner concentrations in 47 clump clusters, 27 chains, and 131 spiral
galaxies were measured from the NICMOS images, and the BViz magnitudes
of the bulges were measured for the same regions using the ACS images
that are degraded to the resolution of NICMOS. The BViz magnitudes of
all the prominent clumps, BLCs, and bulges in 184 clump clusters, 112
chains, and 118 spiral galaxies were also measured from the ACS images.
There are 898, 406, and 845 such objects in the three galaxy types,
respectively.

\begin{figure}\epsscale{1}
\plotone{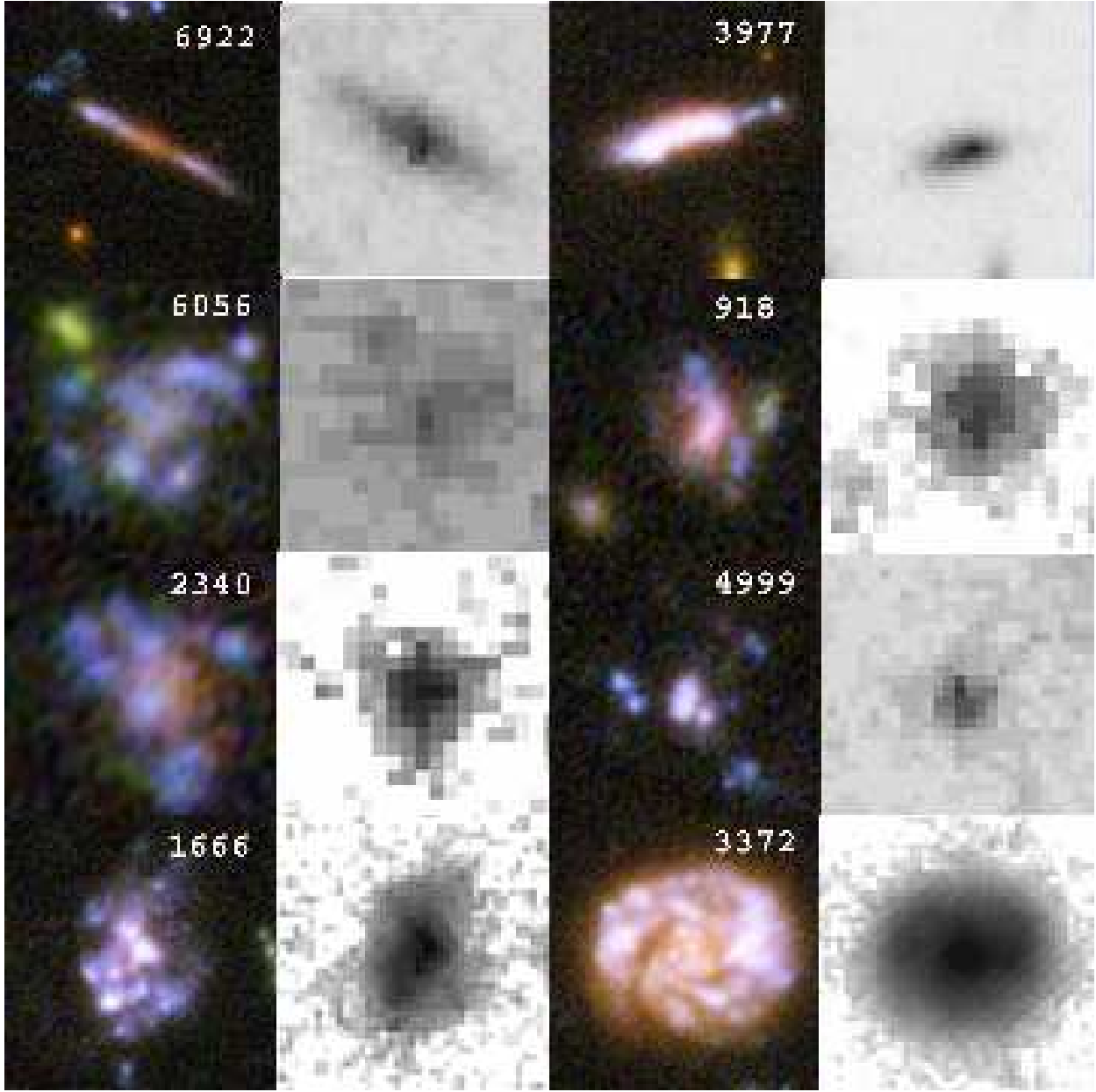} \caption{Color ACS Skywalker images on the left
and NICMOS H-band images on the right for 8 galaxies in the UDF. UDF
numbers are shown, their morphological types are (ch $=$ chain, cc $=$
clump cluster, sp $=$ spiral): UDF 6922 (ch), UDF 3977 (ch), UDF 6056
(cc), UDF 918 (cc), UDF 2340 (cc), UDF 4999 (cc), UDF 1666 (cc), UDF
3372 (sp). Prominent clumps in H-band, which often appear as reddish
objects in the color images, are considered to be bulge-like clumps
(BLCs).  In all of the cases here, BLCs are centralized. (Image
degraded for astroph.) }\label{fig:bulge-jennafig1copy}\end{figure}

Boxes rather than circles were used to define magnitudes because the
clumps are small and pixelated; a typical clump has a diameter of 3 to
5 pixels. No background was subtracted because the surrounding fields
are irregular and because the clumps are much brighter than the
background (see Figs. \ref{fig:bulge-jennafig1copy} -
\ref{fig:bulge-jennafig4b-6108}, discussed below). The sky value in
each image is essentially zero counts. Examples of boxes used for clump
measurements are shown in Figure \ref{fig:UDF2510,1666,3372linear}. In
order from left to right, the galaxies in Figure
\ref{fig:UDF2510,1666,3372linear} correspond to the color images in
Figure \ref{fig:bulge-jennafig4b-6108} (top right), and Figure
\ref{fig:bulge-jennafig1copy} (lower left and lower right, resp.). The
images in Figure \ref{fig:UDF2510,1666,3372linear} are in the $i_{775}$
band and have a grayscale with a linear stretch. For clump clusters,
the clumps are typically more than 3 times brighter than the
surrounding field in $i_{775}$ band. For spirals and chains, the clumps
are 2 to 4 times brighter than the surrounding field in $i_{775}$ band.
These contrasts are larger at shorter wavelengths. The uncertainty in a
given clump brightness can be as much as a factor of two depending on
the box size. Because the same box is used for each passband, the clump
color is much less variable than the clump magnitude. Colors typically
vary by less than 0.05 mag for different box placements or sizes.

Bruzual \& Charlot (2003) model spectra were used to determine model
colors for comparison with the colors of the clumps, BLCs, and bulges.
Problems with TP-AGB stars (e.g., Maraston 2005) affect mostly the
restframe passbands beyond R, which are not generally included in our
ACS measurements for typical redshifts. Such stars could contribute to
the interpretation of NICMOS observations of bulges/BLCs for $z>1.5$,
but comparisons between ACS+NICMOS fits at low angular resolution and
ACS-only fits at high resolution (shown in Figs.
\ref{fig:nicmos_bulgeandclumps_mass_vs_z_alltypes}
-\ref{fig:nicmos_bulgeandclumps_ext_vs_z_alltypes}) do not show
significant trends, suggesting that TP-AGB contamination is less
important than other uncertainties, such as that resulting from low
NICMOS resolution.  For the models, we considered a decaying star
formation history with separate start times and decay rates for each
clump, intervening cosmological Hydrogen absorption (Madau 1995), and
internal dust absorption. A $\Lambda$CDM cosmology was assumed (Spergel
et al. 2003). Photometric redshifts were determined by Elmegreen et al.
(2007a), using the method of Coe et al. (2006). They are accurate to
$\Delta z\sim0.04(1+z)$.

The galaxies identified in the Hubble ACS images were located in the
Spitzer images using the online cutout feature. If a galaxy showed up
in the IRAC images, its integrated light was measured in a 5x5 pixel
box; if it was significantly larger or smaller than this, boxes were
drawn for photometry at the appropriate size. Zeropoints from the
online handbook were used for the conversion of IRAC counts into
magnitudes for comparison with the ACS photometry. Approximately half
of the sample was contaminated in the IRAC images due to blending with
nearby neighbors because of the poor resolution. These cases were not
considered. The Spitzer magnitudes were not used for any derived clump
or bulge properties because these subcomponents were unresolved.

\section{Bulge-like Morphologies in Clumpy Galaxies}
\label{sect:morph}

\begin{figure}\epsscale{1}
\plotone{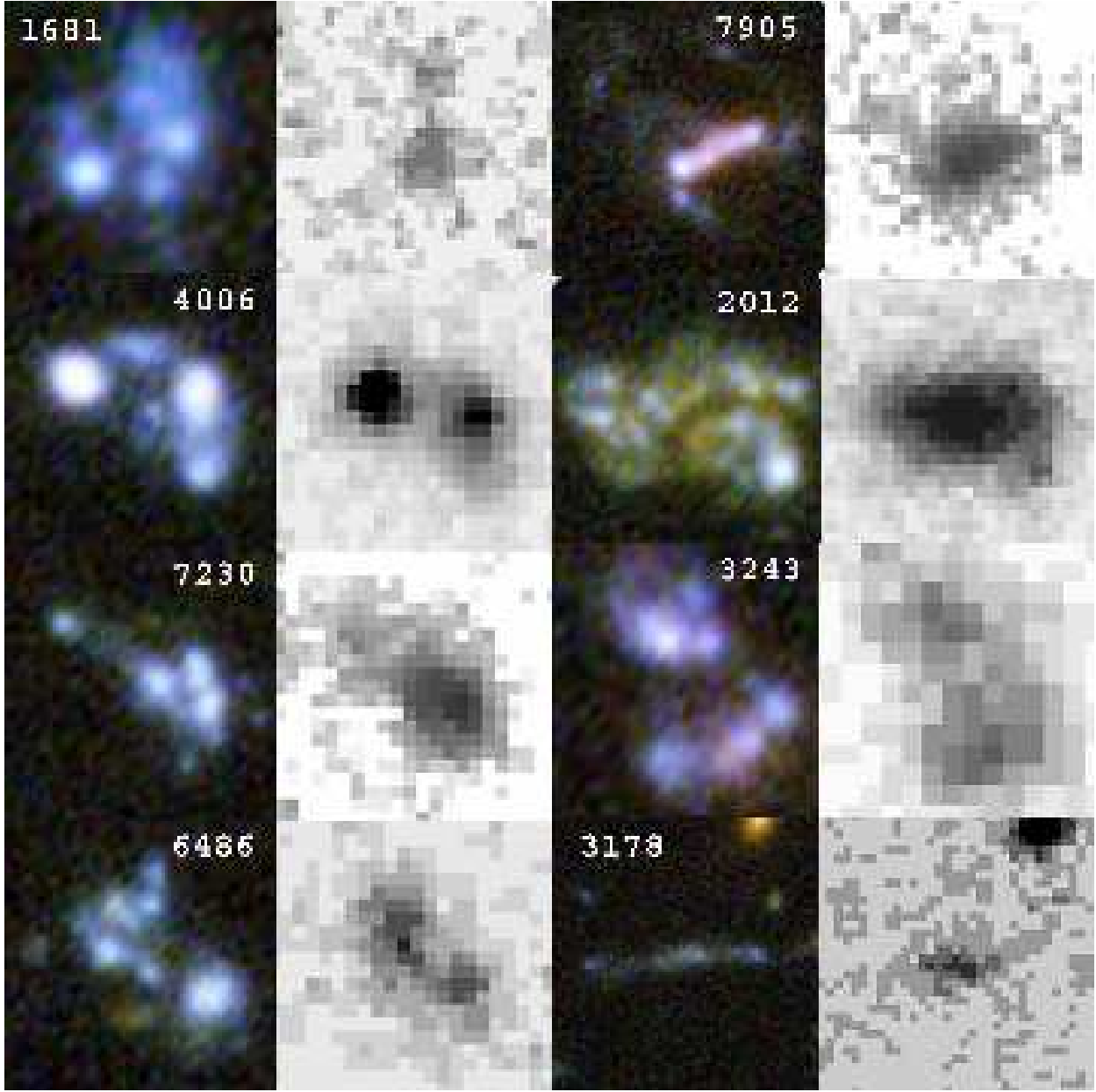} \caption{Color ACS and NICMOS H-band images of
8 galaxies in the UDF without obvious BLCs. Morphological types as in
Fig 1. are: UDF 1681 (cc), UDF 7905 (cc), UDF 4006 (cc), UDF 2012 (cc),
UDF 7230 (cc), UDF 3243 (cc), UDF 6486 (cc), UDF 3178 (ch). The NICMOS
features are relatively large regions of the disks. (Image degraded for
astro-ph.) }\label{fig:bulge-jennafig2new}\end{figure}

\begin{figure}\epsscale{1}
\plotone{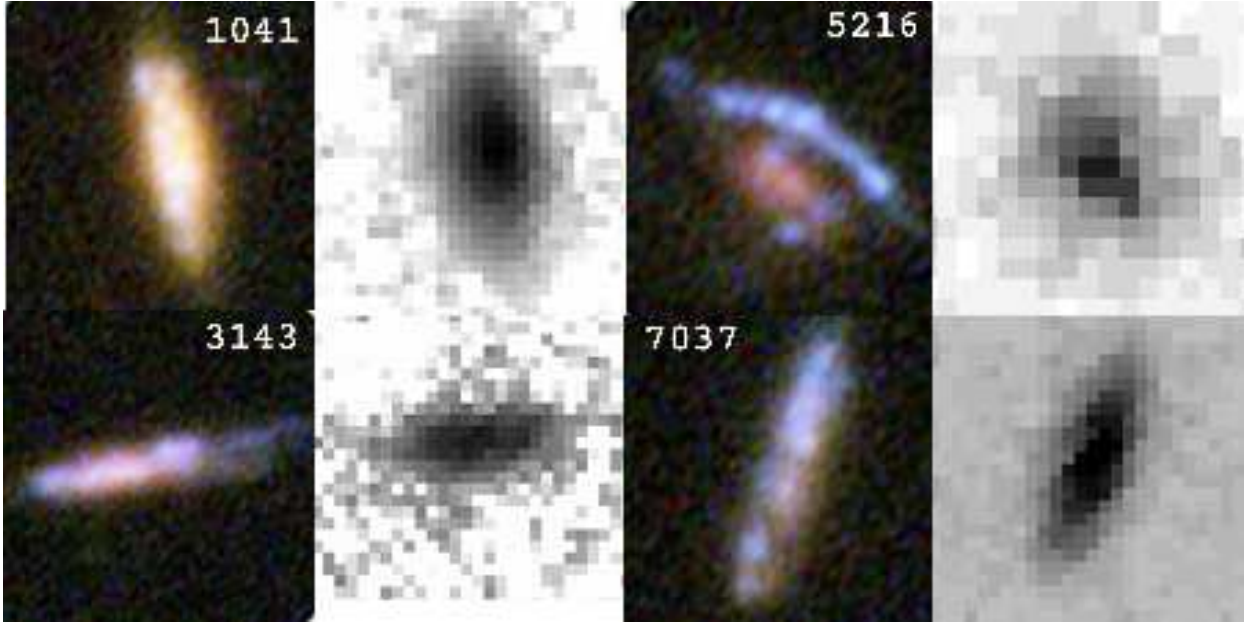} \caption{Color ACS and NICMOS images of 4
galaxies with diffuse K-band emission and no obvious BLCs. The
morphological types are: UDF 1041 (cc), UDF 5216 (cc), UDF 3143 (ch),
UDF 7037 (ch). The blue and red emissions in UDF 5216 could be from two
separate galaxies, but the near-parallel oval alignment of these
features suggests they are in the same inclined disk with the blue part
from star formation in a ring and the red part from a central core. A
more face-on version of this configuration could be something like UGC
2340 in Fig. 1. (Image degraded for astro-ph.)
}\label{fig:bulge-jenna3copy}\end{figure}

\begin{figure}\epsscale{1}
\plotone{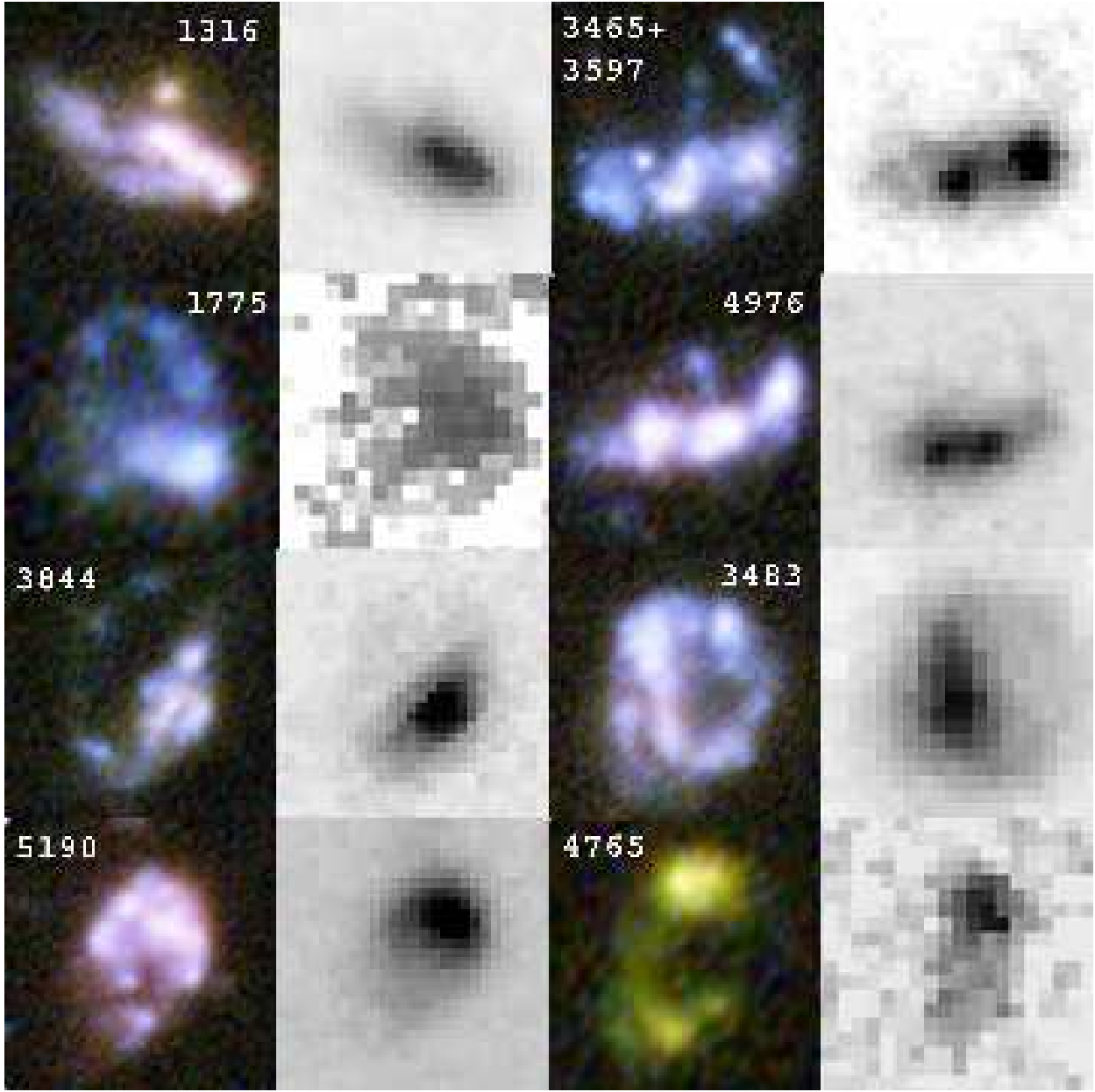} \caption{Color ACS and NICMOS images of
ring-like galaxies, some of which show BLCs where H-band emission is
concentrated into a large mass. The morphologies are: UDF 1316 (cc),
UDF 3465+3597 (cc), UDF 1775 (cc), UDF 4976 (ch), UDF 3844 (cc), UDF
3483 (cc), UDF 5190 (cc), UDF 4765 (cc). Among these, we classify the
two in the top row as bulgeless, the first in the second row (UDF 1775)
as having an offset BLC, the next one, UDF 4976, as having a
centralized BLC, the next three an offset BLC, and the last one, in the
lower right, no BLC. When a large IR concentration is not centralized,
as in many of these case, the bulge-like nature of it is ambiguous.
(Image degraded for astro-ph.)
}\label{fig:bulge-jennafig4acopy}\end{figure}

\begin{figure}\epsscale{1}
\plotone{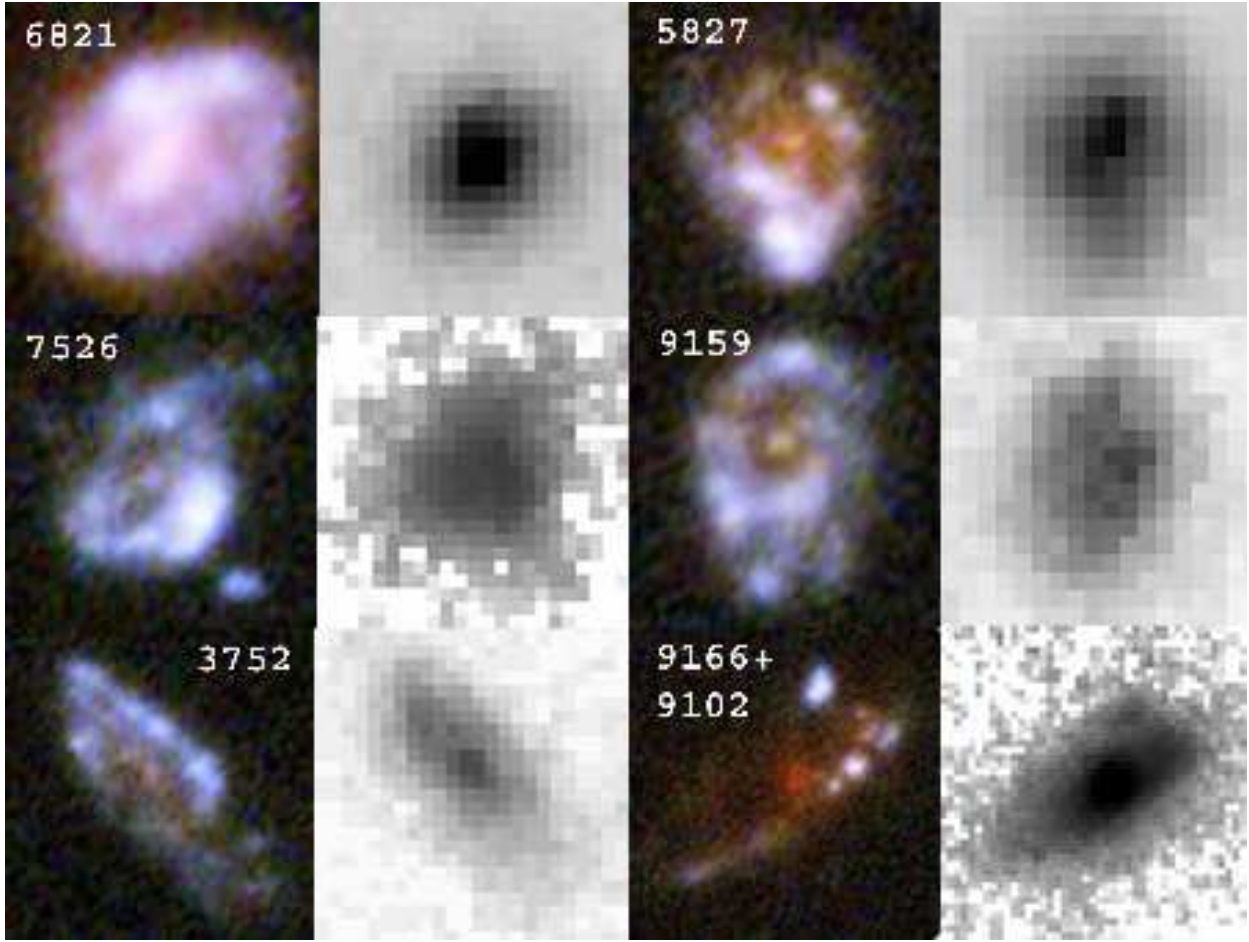} \caption{More ring-like galaxies: UDF 6821
(cc), UDF 5827 (cc), UDF 7526 (cc), UDF 9159 (cc), UDF 3752 (cc), UDF
9166+9102 (cc). We suggest that all of these have centralized bulges or
BLCs except for the first one in the second row, UDF 7526, which has an
offset BLC. For this case there is a faint blue clump near the center
but the H-band emission is concentrated in the ring. (Image degraded
for astro-ph.) }\label{fig:bulge-jennafig4b-6108}\end{figure}

Figures \ref{fig:bulge-jennafig1copy}-\ref{fig:bulge-jennafig4b-6108}
show examples of chains, clump clusters, and spirals having various ACS
and NICMOS BLC or bulge types.  Each galaxy is shown as a pair of
images with the ACS color
Skywalker\footnote{http://www.aip.de/groups/galaxies/sw/udf/index.php}
version on the left and the NICMOS H-band version on the right.  All of
the galaxies considered in this paper are larger than 10 pixels
diameter in the UDF ACS images.  Figure \ref{fig:bulge-jennafig1copy}
has an assortment of chains, clump-clusters and spirals, all with red
centralized BLCs. Prominent blue clumps, presumably from star
formation, surround the BLCs. Figure \ref{fig:bulge-jennafig2new} has
galaxies with no obvious bulges but irregular clumpy structures that
are similar in visible and NICMOS bands. Figure
\ref{fig:bulge-jenna3copy} has galaxies with bright NICMOS emission in
a diffuse form, like a whole disk, and not clumped like a bulge.
Figures \ref{fig:bulge-jennafig4acopy} and
\ref{fig:bulge-jennafig4b-6108} have ring-like galaxies which may or
may not have a bright NICMOS clump reminiscent of a bulge (bulge
identifications are in the figure caption; for more on ring galaxies at
high redshift, see Elmegreen et al. 2007b).

\section{Clump Colors and Magnitudes}
\label{sect:color}

The NICMOS online catalog gives integrated BVizJH magnitudes for each
galaxy. These were used in the top panels of Figure
\ref{fig:nicmos_I-ZvsZwithclumps} to plot (i$_{775}$-z$_{850}$) versus
redshift for whole clump cluster, chain, and spiral galaxies.  The
middle panels show these colors for the bulge or bulge-like regions of
each type. The pluses represent centralized bulges or BLCs and the
circles are significantly off-center BLCs. The small blue dots in the
middle row of panels are NICMOS observations of bright central regions
that are diffuse like a disk, and not clumpy like a bulge (these are
considered to be non-bulge galaxies, along with other cases where there
are neither diffuse nor clumpy bright NICMOS features).  Blue symbols
in this middle row (pluses, circles, dots) use the online ACS images
that are blurred by a factor of 3 to match the NICMOS resolution. The
red crosses in the middle row of panels are the colors for the red
clumps in the ACS images, presumed to be BLCs because they are at the
same positions as the NICMOS BLCs; they were measured with the full
resolution of the ACS whenever they could be detected there.  The
bottom panels show the colors for each prominent clump, measured in the
high-resolution ACS image, whether or not it was identified with a
bulge or BLC.

Each panel in Figure \ref{fig:nicmos_I-ZvsZwithclumps} has a peak near
$z=1.3$, where i$_{775}$ and z$_{850}$ correspond to restframe
wavelengths of 336nm and 370nm. These approximately straddle the Balmer
jump at $\sim370$nm, so this peak is from that jump. Evidently the
bulge/BLC population is young enough to have a Balmer jump, but not as
young as the clumps, which are typically bluer than the bulges/BLCs and
the whole galaxies.  This relative clump blueness is also evident from
the color images, although some bright central regions, presumably
bulges, are as blue as the clumps in the rest of those galaxies.
Comparison with the evolutionary models in Conselice (2008) shows that
the bulges and BLCs have the same $i_{775}-z_{850}$ colors as
integrated Scd galaxies.

We compared our UDF bulges to those measured in the Hubble Deep Fields
North and South by Ellis, Abraham, \& Dickinson (2001) out to
$z\sim1.5$. They showed that bulges in early and late type spirals are
significantly bluer than the centers of elliptical and S0 galaxies at a
given redshift, as measured by their V$_{606}$-I$_{814}$ colors. Ellis
et al. also included NICMOS measurements of their galaxy centers and
integrated galaxies. In order to compare our results for bulges in UDF
clump cluster and chain galaxies with their HDF measurements, we needed
to transform their Vega mag to AB mag by applying the zeropoints in
Lucas et al. (2003), which give $(J-H)_{AB}=(J-H)_{Vega}-0.59$. We
selected their non-peculiar, non-blended galaxies, which they
classified automatically based on central concentration and asymmetry
as well as visually. At redshifts out to $z\sim1.5$, which encompasses
the bulk of the Ellis et al. sample, our clumpy galaxy bulge-like
objects have about the same or slightly bluer colors as their HDF
intermediate type spiral bulges. The integrated galaxy colors are also
correspondingly slightly bluer for the clumpy galaxies than for the HDF
elliptical and spiral galaxies.

The apparent $z_{850}$ magnitudes for the same galaxies, bulges or
BLCs, and clumps are shown versus redshift in Figure
\ref{fig:nicmos_appmag_vs_z}, using the same symbol notation. A fixed
magnitude of 26 is shown in each panel as a dotted red line to
facilitate comparisons; the clumps are typically 1 to 2 magnitudes
fainter than the bulges/BLCs, which are 2 to 3 magnitudes fainter than
the whole galaxies. At high redshift, the bulges in spirals are
somewhat brighter in $z_{850}$ than the BLCs in clump clusters and
chains (the same is true for H band, discussed below).

In the top right panel of Figure \ref{fig:nicmos_appmag_vs_z}, the
distance modulus is plotted as a curve with its scale on the right. The
absolute magnitude is equal to the difference between the apparent
magnitude of the plotted point and the distance modulus at the same
redshift. Absolute magnitudes of $-18$ mag and $-20$ mag are indicated
by dotted lines in the lower two rows. These values bracket the range
of absolute magnitudes for the brightest BLCs and clumps in clump
clusters and chains. Bulge/BLC $z_{850}$ magnitudes increase slightly
with redshift, following the green dashed lines. This is also observed
in H-band (Fig. \ref{fig:nicmos_appmagH_vs_z} below). The bulge/BLC
masses are relatively constant with $z$ (Sect. \ref{sect:results}), so
the change in $z_{850}$ is probably a bandshifting effect; i.e., the
restframe blue colors of the bulges/BLCs are fainter than the restframe
red colors.

The distribution of clumps in the bottom row of Figure
\ref{fig:nicmos_appmag_vs_z} is more horizontal; i.e., the apparent
magnitude is about constant with redshift for clumps. This is partly a
selection effect because star-forming objects as faint as those seen
locally cannot be observed at higher redshifts. It is also partly a
bandshifting effect: typical clump masses and ages are about constant
with redshift, so the brightening in the observed $z_{850}$ band is the
result of an increasing shift of blue emission from intense star
formation into this filter with higher redshift. This is opposite to
the shift for bulges/BLCs because of the intrinsic color difference.

Restframe apparent magnitudes, $B_{435}$, were determined for each
galaxy, bulge, BLC, and clump by interpolation between the four ACS
passbands and plotted in the same way as in Figure
\ref{fig:nicmos_appmag_vs_z}. The restframe distributions were very
similar to the observer-frame distributions in Figure
\ref{fig:nicmos_appmag_vs_z}, so the results are not shown here. The
similarity arises because the magnitude range for all the points (from
the diverse galaxy brightnesses) is much larger than the magnitude
range for any one galaxy among the different passbands.

The restframe absolute magnitudes, $M_{B435}$, are shown in Figure
\ref{fig:nicmos_absmag_vs_z} using the same symbol notation as in
Figure \ref{fig:nicmos_appmag_vs_z}. The constant value of $-16$ mag is
drawn for comparison; this value is typically considered to be a
dividing point between local dwarf and spiral galaxies. Evidently, most
clumps at intermediate to high redshifts are brighter than local dwarf
galaxies. For the determination of restframe magnitudes, the high
redshift points have no ACS observations for interpolation, so the ACS
$z_{850}$ absolute magnitude is plotted instead, as an estimate.

Figure \ref{fig:nicmos_J-HvsZ} shows NICMOS J-H colors versus redshift
for whole galaxies (top) and bulges or BLCs (bottom) using the NICMOS
catalog in the first case and IRAF measurements on the NICMOS images in
the second case. Infrared colors are not available for the clumps
because they are usually not present in NICMOS images or they are not
resolved by NICMOS. The J-H colors straddle the Balmer jump at a
redshift of 2.6 but no features are observed in the figure there.

Figure \ref{fig:nicmos_appmagH_vs_z} shows the NICMOS apparent H-band
magnitudes versus redshift for whole galaxies (top) and bulges
(middle). A fiducial magnitude of 26 is shown as a dotted red line to
aid comparisons. The apparent magnitudes of the bulges and BLCs tend to
increase with redshift parallel to the dashed line of constant absolute
magnitude; this is probably a result of bandshifting because the
bulge/BLC masses do not vary like this (Sect. \ref{sect:results}) and
because the IR colors, $i_{775}-z_{850}$, $J-H$, etc., are positive by
several tenths each. Thus higher redshifts show more restframe blue
light in H-band, and this is fainter than the restframe red light for
the bulges. A similar increase in $z_{850}$ magnitude with redshift was
seen in Figure \ref{fig:nicmos_appmag_vs_z}.

Figure \ref{fig:nicmos_appmagH_vs_z} indicates that spiral bulges at
high redshift are $\sim2$ mag brighter than the BLCs in clump cluster
and chain galaxies at high redshift, just as whole spiral galaxies are
brighter in H-band than the others. The relative brightness of spiral
bulges (also in $z_{850}$ band, shown above) is explained in Section
\ref{sect:results} to be a result of their larger bulge masses compared
to clump cluster and chain BLC masses.  The bottom row in Figure
\ref{fig:nicmos_appmagH_vs_z} shows the differences between the
bulge/BLC and the galaxy H-band magnitudes. The average values of these
differences are shown by dotted red lines. For clump clusters, chains,
and spirals, the average magnitude differences are $2.58\pm0.75$,
$3.09\pm1.03$, and $2.62\pm0.75$, respectively. To the extent that the
H band is a reasonable tracer of stellar mass, these magnitude
differences correspond to a relative bulge/BLC mass compared to the
galaxy mass of $\sim1.7\times10^{-3}$.

Figure \ref{fig:nicmos_spitzer} shows the ($3.6\mu m - 4.5 \mu m$)
colors as a function of redshift for the clump cluster and chain
galaxies, sorted by bulge type as before, and also for spiral galaxies
in the UDF for comparison. There is no clear difference among the
different types of galaxies, although most of the galaxies with bulges
or BLCs are redder than non-bulge galaxies.

\section{Age and Mass Models}
\label{sect:models}

\subsection{Method}
\label{sect:method}
 The photometric colors were compared with colors
obtained from redshifted low-resolution spectral templates of stellar
populations with a Chabrier initial mass function, as tabulated by
Bruzual \& Charlot (2003). We confined our detailed study to the
Bruzual \& Charlot models with metallicities of 0.008 (equal to 0.4
solar) because metallicities for these types of objects are not
commonly observed yet and this level of metallicity seems reasonable
for partially developed galaxies. We note that the metallicity
determined for the clump cluster galaxy UDF~6462 from high resolution
spectral observations is $\sim0.5$ solar on average, and about 20\%
higher than this in a reddish object that resembles a bulge (Bournaud
et al. 2008).  The value we choose is comparable to this measurement.
Results for lower metallicities of 0.004 and 0.0004 will be discussed
in Section \ref{sect:results}. Corrections to the model spectra were
made to account for intervening Hydrogen absorption, taking the Lyman
series up to order 20 and including continuum (Madau 1995), and for
dust in the high redshift galaxy. Photometric redshifts are from
Elmegreen et al. (2007a).

The procedure for fitting the stellar population was as follows (see
also Elmegreen \& Elmegreen 2005).  We considered three parameters for
each clump and bulge/BLC fit: the starting time of star formation in
that clump or bulge/BLC (i.e., the ``age,'' $T$, of the region in the
galaxy's restframe), the exponential decay rate of star formation in
that region ($\tau$), and the extinction in that region. The possible
starting times ranged from zero in the galaxy's restframe, back to the
relative time corresponding to a redshift of $z=10$. The decay times
were $\tau=10^7,\;3\times10^7,\;10^8,\;
3\times10^8,\;10^9,\;3\times10^9$ and $10^{10}$ years. With these two
parameters, the star formation history in the clump under consideration
was assumed to begin at a time in the past equal to the ``age'' and
then decay exponentially over time with a timescale $\tau$ until the
current restframe time of the galaxy. Dust absorption was included as a
third parameter of the fit using the wavelength dependence in Calzetti
et al. (2000) with the short-wavelength modification in Leitherer et
al. (2002), and using discrete multiples of $M_A$ times the redshift
dependence for galaxy intrinsic $A_V$ in Rowan-Robinson (2003). These
multiples were $M_A=0.25$, 0.5, 1, 2, 4, and 8 times the Rowan-Robinson
extinctions for the redshift of each galaxy. For each fit of these
three parameters, there are 4 passbands and 3 colors from the ACS
observations, and in the case of bulges or BLCs, there are two more
colors from NICMOS. Thus the fits are mathematically well determined,
although the uncertainties can be large for individual values,
particularly for age and extinction. We discuss the fit uncertainties
for clump masses and ages in Section \ref{sect:results}. For clumps or
bulges/BLCs with ACS observations, 4 passbands were fit to the models.
For bulges/BLCs with NICMOS observations in $J$ and $H$ bands, six
passbands were fit to the models, two from NICMOS and four more from
the $3\times$ blurred ACS images that match the resolution of NICMOS.

We note that the colors of these galaxies were also used to determine
photometric redshifts, and along with the redshifts, galactic spectral
types $t_b$ (Elmegreen 2007a). The colors used for redshifts are for
whole galaxies and not for clumps, while the new fits given here are
for the clump colors alone. Thus the same observations are not used for
both redshift and clump properties, and the clump 3-parameter fits, for
$(T,\tau,M_A)$, are not degenerate with the fit for redshift. For
example, we find young clumps in spirals with a wide range of
redshifts, even though spirals tend to have low $t_b$ (old average
ages) at low redshift (Elmegreen 2007a).

For each age $T$ and decay time $\tau$, we determined the model
spectrum of the stellar population by integrating over the product of
the time-dependent spectra from tables in Bruzual \& Charlot and the
time-dependent star formation rate, $\exp\left(-[T-t]/\tau\right)dt$,
for time $t$ in the range $(0,T)$. These tables are normalized to 1
$M_\odot$ of stars for each spectrum, so we also integrated over the
product of the star formation rate and the time-dependent remaining
stellar mass (the remaining stellar mass is what remains from the
original 1 $M_\odot$ after stellar evolution for the time $T-t$). We
then corrected the integrated model for intrinsic dust absorption,
assuming one of the absorption multipliers, $M_A$, redshifted the
corrected spectrum to the redshift of the galaxy, and corrected it
again for intervening Hydrogen absorption lines.  This redshifted
population spectrum was then multiplied by the normalized, HST
filter/throughput functions (downloaded from the HST website) to get
the model absolute magnitudes per unit solar mass of stars formed, and
from these, the model colors for the assumed $(T,\tau,M_A)$
combination. The absolute AB magnitude in $i_{775}$ band was converted
to an apparent magnitude using the distance modulus at the galaxy
redshift for a $\Lambda$CDM cosmology (Spergel et al. 2003; Carroll,
Press, \& Turner 1992), and the ratio of the observed apparent flux to
the model apparent flux in the $i_{775}$ band was multiplied by the
model mass to get the total stellar mass for that $(T,\tau,M_A)$
combination.

To determine how well each $(T,\tau,M_A)$ combination fits the observed
clump or bulge/BLC, the rms differences between the model colors and
the observed colors were determined for each $(T,\tau)$ at a particular
$M_A$. These rms differences were binned into groups with values
incremented by 0.1. We want to consider only models with the lowest rms
deviations from the observations, but there are usually many models
with comparably good fits. To find the best of these, we considered
those in the two bins with the lowest rms values for that clump, still
at a fixed $M_A$, and then picked the one with the largest number of
entries. The preferred values for $T$, $\tau$, and stellar mass at that
$M_A$ were taken to be the averages of all those contributing to this
most populated low-rms bin, where the average was weighted by the
inverse square of the rms deviation. The six different dust
multipliers, $M_A$, were assessed in a different way. Each gave the
preferred rms fit just described, but some had lower preferred rms
values than others.  Among all those with the same lowest rms value for
each clump, we averaged together the $T$, $\tau$, mass, and extinction
values with weighting factors equal to the number of entries in the
selected rms bin from the $(T,\tau)$ fits.  Thus the final solution to
these quantities is a weighted average of the best individual fits (and
a corresponding rms around that average). Note that only $T$, $\tau$
and extinction need to be fitted to the three observed ACS colors; the
mass follows from any one magnitude, which is an independent
observation; in practice we used the $i_{775}$ magnitude for mass
because that has the lowest statistical error. Many other methods for
optimizing the models were also attempted, but the one just described
consistently gave good fits to the observed spectral energy
distributions (SED) of the clumps and bulges.  All of the other methods
gave about the same individual clump masses, which are somewhat
invariant to compensating changes in $T$, $\tau$ and $M_A$, but the
clump ages depended on the method somewhat, with $\sim50$\% variations
typical among the different methods (see discussion of uncertainties
below).

\subsection{Results}
\label{sect:results}

Figure \ref{fig:nicmos_colors_2601_3209} shows the SEDs and model fits
for two galaxies, UDF 4999 and UDF 7230, which were also shown in
Figures \ref{fig:bulge-jennafig1copy} and \ref{fig:bulge-jennafig2new}
as examples of clump clusters with and without BLCs, respectively. In
Figure \ref{fig:nicmos_colors_2601_3209}, the observations are shown by
the symbols and the best-fitting models are shown by the lines. Values
of extinction, age, mass, and restframe absolute magnitude $B_{435}$
are given for each line. The SED fits were viewed in this way for all
of the galaxies in Figures \ref{fig:bulge-jennafig1copy} and
\ref{fig:bulge-jennafig2new}; they were all reasonably good fits like
these (the Fig. \ref{fig:nicmos_colors_2601_3209} galaxies were chosen
because they have a good number of clumps and the fitting curves are
spread out enough to see well on the plot.)

Each SED model for a bulge, BLC, or clump is a $(T,\tau,M_A)$
combination which has model colors that differ from the observed colors
by a small amount. The deviation, $\Delta {\rm color}$, is calculated
in quadrature as
\begin{equation}
\Delta {\rm color}^2 = \left(\Delta [B-V]\right)^2+\left(\Delta
[V-i]\right)^2+\left(\Delta [i-z]\right)^2+\left(\Delta [z-J]\right)^2
+\left(\Delta [J-H]\right)^2 \end{equation} where
\begin{equation}\Delta(B-V)=
\left(B_{435}-V_{606}\right)_{\rm
model}-\left(B_{435}-V_{606}\right)_{obs.},
\end{equation}
and so on for the other colors.  For the pure-ACS observations, the sum
only extends to the $i_{775}-z_{850}$ color. As mentioned in the
previous section, the final fit is determined by averaging over the
masses, ages, decay times, and extinctions among those models which
give the lowest $\Delta {\rm color}$, when this $\Delta {\rm color}$ is
binned in intervals of 0.1. Histograms of these lowest $\Delta {\rm
color}$ values, measured in magnitudes, are shown in Figure
\ref{fig:nicmos_plotrms} for the clumps on the left and the bulges or
BLCs on the right.  The histograms indicate that a typical fit is good
to between 0.1 and 0.3 of a magnitude in the quadrature summed color,
on average, which corresponds to $\sim0.06$ to $\sim0.2$ mag in each
color for 3 colors. This is in agreement with the differences between
the lines and the plotted points in Figure
\ref{fig:nicmos_colors_2601_3209}. Recall that the measurement error
for each color is about 0.05 magnitude, depending on clump brightness,
so the SED fits are typically about as good as the measurement errors.

The masses of the bulges, BLCs, and clumps are shown as functions of
redshift in Figure \ref{fig:nicmos_bulgeandclumps_mass_vs_z_alltypes};
the ages are shown in Figure
\ref{fig:nicmos_bulgeandclumps_age_vs_z_alltypes}, and the restframe
V-band extinctions are in Figure
\ref{fig:nicmos_bulgeandclumps_ext_vs_z_alltypes}. Each clump, BLC, or
bulge corresponds to one point on each figure, which is the best
simultaneous fit to all three quantities using the three or more
colors, as described above. The symbol types are the same as before: in
the top rows, the blue plus signs are centralized bulges/BLCs measured
with NICMOS and the four ACS bands using NICMOS resolution, the blue
circles are off-center BLCs measured the same way, and the blue dots
are non-bulge concentrations using NICMOS and low-resolution ACS
images. The red crosses are bulge/BLC measurements using only the ACS
full resolution passbands. The second row in each figure shows the
clump properties for clumps that are not bulges or BLCs, all modeled
with only the four high-resolution ACS bands. The dotted red lines
guide the eye to a standard value for ease in comparison between clumps
and bulges. In all cases, the plotted results are weighted averages
from best-fitting models that minimized rms deviations from the
observations, considering simultaneous variations in age, exponential
decay time for the star formation history, and internal extinction, as
discussed in Section \ref{sect:method}.

Before discussing the results in Figures
\ref{fig:nicmos_bulgeandclumps_mass_vs_z_alltypes} -
\ref{fig:nicmos_bulgeandclumps_ext_vs_z_alltypes}, we consider the
uncertainties resulting from model fitting, age and extinction
degeneracy, and metallicity. Figure
\ref{fig:nicmos_bulgeandclumps_age_vs_z_alltypes_rms} shows the rms
deviations of the mass determinations, in the top two panels, and the
age determinations, in the bottom two panels, for each fit in Figures
\ref{fig:nicmos_bulgeandclumps_mass_vs_z_alltypes} and
\ref{fig:nicmos_bulgeandclumps_age_vs_z_alltypes}, considering all of
the individual model results that share the same low value of $\Delta
{\rm color}$. Recall that the fits were determined from averages among
all of these low-$\Delta {\rm color}$ models, so the plotted points in
Figure \ref{fig:nicmos_bulgeandclumps_age_vs_z_alltypes_rms} are the
rms values around those averages. They could have been included in
Figures \ref{fig:nicmos_bulgeandclumps_mass_vs_z_alltypes} -
\ref{fig:nicmos_bulgeandclumps_age_vs_z_alltypes} as error bars, but
then crowding would make the figures unreadable. A typical rms value
for the bulge mass is 0.15 in the log, which means that the bulge
masses plotted in Figure
\ref{fig:nicmos_bulgeandclumps_mass_vs_z_alltypes} have deviations
among all of the low-$\Delta {\rm color}$ models of a factor of 1.4.
For clumps, $\Delta \log {\rm Mass}\sim0.2-0.3$ for mass, so the error
factor is 1.6 to 2. The error factors are slightly larger for age:
$\Delta \log {\rm age}\sim0.3$ for bulges or BLCs and $\sim0.4$ for
clumps, corresponding to factors of 2 to 2.5, respectively.

Figure \ref{fig:nicmos_clump_massM42_vs_massM52_andage2} shows three
comparisons for age and mass of clumps in clump cluster galaxies. On
the left is a comparison between fits for an extinction equal to the
Rowan-Robinson (2003) value at that redshift and an extinction equal to
twice the Rowan-Robinson value. In the bottom left frame, the mass from
the $M_A=2$ fit is plotted on the ordinate and the mass from the
$M_A=1$ fit is plotted on the abscissa. The white dashed line
represents equality for the two fits. In the top left frame, the age
from the $M_A=2$ fit is plotted versus the age from the $M_A=1$ fit.
The masses hardly change but the ages are a factor of up to $\sim2$
smaller for the higher extinction. This change in age is the result of
a need for an intrinsically bluer star formation region to compensate
for the greater reddening from extinction. The mass does not change
much because the younger age corresponds to a greater light-to-mass
ratio and this greater light compensates for the greater extinction at
the same mass.

In the middle of Figure
\ref{fig:nicmos_clump_massM42_vs_massM52_andage2} is a similar age and
mass comparison for two different metallicities. The abscissa has
values for $Z=0.008$, as assumed throughout this paper (and equal to
0.4 solar), and the ordinate has $Z=0.004$. On the right in the figure,
the abscissa has the same $Z=0.008$, but the ordinate has a much
smaller metallicity, $Z=0.0004$. The fitted clump ages change only
weakly with metallicity, increasing by $\sim50$\% for the lowest $Z$.
This increase is probably the result of lower line blanketing in low
metallicity stars, which gives them a bluer color that has to be
compensated by a larger age to get the same observed color. The fitted
masses increase by $\sim50$\% too at the lowest metallicity, to
compensate for the older ages which have a slightly lower light-to-mass
ratio.

With these uncertainties in mind, we can now comment on the results in
Figures \ref{fig:nicmos_bulgeandclumps_mass_vs_z_alltypes} -
\ref{fig:nicmos_bulgeandclumps_ext_vs_z_alltypes}. Figure
\ref{fig:nicmos_bulgeandclumps_mass_vs_z_alltypes} indicates that the
bulge/BLC mass is about constant between $\sim10^8$ and $10^9\;M_\odot$
for all galaxy types at $z>0.5$. The bulges in spirals are slightly
more massive than the BLCs in the other types at high redshift. The
clump mass is also about constant for clump cluster and chain galaxies
beyond $z\sim0.5$. The bottom row in the Figure shows the ratio of the
clump mass to the bulge/BLC mass for each clump in a galaxy that has a
bulge/BLC measured with the ACS (i.e., using only the ACS measurements
for both). The average of the log of the ratio of the clump to the
bulge/BLC mass is $-0.73\pm0.60$ for clump clusters, $-0.38\pm0.49$ for
chains, and $-1.42\pm1.05$ for spiral galaxies. These average ratios
indicate that clumps in spiral galaxies are smaller compared to their
bulges than the clumps in clump cluster and chain galaxies are to their
BLCs. The low redshift spirals also have lower mass clumps compared to
their bulges than high redshift spirals. In all cases, there appears to
be a general decrease in clump-to-bulge mass ratio with galaxy
maturity. In the middle right panel, the spiral clump masses appear to
increase more systematically with $z$ than the clump cluster and chain
masses, but this increase is not evident in the ratio of the clump to
the bulge/BLC mass. In general, the small masses at $z\lesssim0.5$ are
the result of a selection effect because the field of view for the UDF
has a smaller volume at lower redshift, thereby losing the rare high
mass galaxies, and because the field was chosen to avoid bright nearby
galaxies.

The average ratio of the clump mass to the galaxy mass can be estimated
from the ratios between the H-band bulge/BLC and galaxy fluxes (Fig.
\ref{fig:nicmos_appmagH_vs_z}), which are approximately their ratios of
masses, and from the ratios of clump to bulge/BLC masses obtained from
the models.  The resulting logarithms of the clump to galaxy mass
ratios are $-1.75\pm0.64$ for clump clusters, $-1.62\pm0.49$ for
chains, and $-2.47\pm1.05$ for spirals. As percentages, these ratios
are 1.8\%, 2.4\%, and 0.34\% for the three types. Evidently the clumps
in clump cluster and chain galaxies each contain an average of
$\sim2$\% of their galaxy mass, while each spiral clump contains only
0.3\% of its galaxy mass.

Figure \ref{fig:nicmos_bulgeandclumps_age_vs_z_alltypes} suggests that
bulge/BLC and clump ages decrease slightly for galaxies with increasing
redshifts, particularly for the clumps. Recall that the ages are not as
well determined as the masses because of the age/extinction degeneracy.
Nevertheless, the age-redshift trend is reasonable considering that
increasing $z$ corresponds to generally younger galaxies for which
there is a decreasing available time for star formation to have
occurred.  Also plotted in the figure is the age of the universe as a
function of redshift, assuming the standard model (Spergel et al.
2003). The bulge/BLC ages decrease with $z$ in a manner somewhat
parallel to the decrease in the age of the universe. Bulges/BLCs are
younger than the universe by 2 to 4 Gyr, with the lowest differences
arising for the highest redshifts. The clump ages are generally younger
than the bulge/BLC ages (as expected because the bulges and BLCs are
partially selected on the basis of their reddish color), and they also
show a decreasing trend with increasing $z$, but the scatter is much
larger than the trend. The downward sloping streaks in the clump plots
illustrate the lowest ages for which there is any sensitivity.

Figure \ref{fig:nicmos_bulgeandclumps_ext_vs_z_alltypes} shows
restframe V-band extinction versus redshift for each clump, BLC, and
bulge. The extinction to clumps systematically decreases with redshift.
It can be as high as 3 or 4 magnitudes for low redshift, but the
highest values we fit at high redshift are only about one magnitude.
There is not much difference in the extinction among the three galaxy
types, nor among the bulges, BLCs, and clumps. This extinction trend
with redshift is consistent with expectations for decreasing
metallicity, and also with the decrease in galaxy-wide extinction found
by Rowan-Robinson (2003) and others. Recall that these Rowan-Robinson
extinctions are used as fiducial values for the fits because they are
expected to reflect the general trends with redshift, but the actual
fit does not use the Rowan-Robinson extinction, it chooses between
various multiples ($M_A=0.25$ to 8) times this extinction and then
picks the multiple, and therefore the extinction, giving the lowest rms
deviation from the SED (Sect. \ref{sect:models}).

An interesting implication of the trend in Figure
\ref{fig:nicmos_bulgeandclumps_ext_vs_z_alltypes} is that it has the
same sense as the trend in Figure
\ref{fig:nicmos_bulgeandclumps_age_vs_z_alltypes}: both decrease with
redshift. The age-extinction degeneracy has the opposite trend. For a
given clump, a higher value of extinction leads to a fit with a lower
value of age (Fig. \ref{fig:nicmos_clump_massM42_vs_massM52_andage2}).
Thus the decreasing fitted age with redshift cannot be the result of an
increasing fitted extinction with redshift.  Also, the decrease in
extinction with age implies that the uncertainty in age from the
age-extinction degeneracy decreases with redshift.

Figure \ref{fig:nicmos_bulgeclump_histograms} plots histograms of the
ratios of the clump and bulge/BLC properties determined by the models:
mass, age, exponential decay time of the star formation rate, and
extinction. The vertical dotted line in each panel indicates a zero
value for the log, which means that the clump and bulge/BLC properties
are the same. The mass ratios were already shown on a clump-by-clump
basis in the bottom panels of Figure
\ref{fig:nicmos_bulgeandclumps_mass_vs_z_alltypes}. The trend noted
before, that the clumps in spirals are slightly smaller relative to
their bulges than the clumps in clump clusters and chain galaxies, is
seen again in Figure \ref{fig:nicmos_bulgeclump_histograms} as a
systematic shift leftward for the spiral galaxy mass ratio distribution
function. The spiral clumps are also slightly younger compared to their
bulges than the clump cluster and chain galaxy clumps.  The other
panels in Figure \ref{fig:nicmos_bulgeclump_histograms} show no
systematic shifts between bulges/BLCs and clumps for different galaxy
types; i.e., neither for their star formation decay times nor for their
extinctions.

A Kolmogorov-Smirnov test was applied to test the similarity of the
distributions in $\log M_{clump}/M_{bulge}$ for spirals, clump clusters
and chains. There are 726, 150, and 12 clumps in these three
histograms, respectively.  The hypothesis that the clump cluster and
spiral mass fraction distributions are drawn from the same population
is rejected at very high confidence level, $1-2\times10^{-16}$; for
spirals and chains, the hypothesis is rejected at a confidence level of
$1-2\times10^{-4}$.  Thus the clump-to-bulge mass ratios are
intrinsically smaller for spirals than for the other two types. For the
age histograms, the rejection probabilities are $1-10^{-7}$ and 0.996
for clump clusters and chains, respectively, compared to spirals. Thus,
the ages of the clumps in clump clusters and spirals are intrinsically
closer to the ages of the bulge-like objects in those galaxies (in the
cases where there are bulge-like objects), than the ages of clumps in
spirals are to the ages of the bulges in spirals. Clump cluster and
chain BLCs, when they exist, are more similar to the star forming
clumps in these galaxies than spiral bulges are to the star forming
clumps in spirals. This is the main result of this paper.

The averages and dispersions of the histograms in Figure
\ref{fig:nicmos_bulgeclump_histograms} do not change much for different
metallicities. All of the fits discussed in this paper assume
$Z=0.008$, which is 0.4 solar, but the histogram trends are still
present if both the clumps and the bulges have $Z=0.004$, and they are
also present if the clumps have $Z=0.008$ and the bulges have
$Z=0.004$, and vice versa.  Table 1 gives average values and standard
deviations of the distribution functions for clump-to-bulge ratios for
mass, age, decay time, and extinction.  For example, in Figure
\ref{fig:nicmos_bulgeclump_histograms}, the average value of
$\log\left(M_{clump}/M_{bulge}\right)$ for spirals is $-1.42$ and the
standard deviation of the distribution is 1.05; we refer to combination
as $-1.42\pm1.05$ in the table. The tabulated values hardly change for
the different metallicities.

\section{Discussion}
\label{sect:disc}

Clump cluster and chain galaxies are dominated by giant star-forming
regions in the ACS images. Their spectral types, $t_b$, are generally
large, $t_b\gtrsim5$, indicative of extreme starbursts ($t_b$ were
determined in Elmegreen et al. 2007a as part of the fit to photometric
redshift). These objects also emit at relatively long restframe optical
wavelengths, observed here as redshifted emission in the NICMOS and
IRAC bands. Because the $J-H$ and $3.6\mu m-4.5\mu m$ colors are fairly
uniform among clump clusters and chains, and also comparable to the
colors of spirals (Figs. \ref{fig:nicmos_J-HvsZ} and
\ref{fig:nicmos_spitzer}), most clump clusters and chains cannot be
random assemblies of young star formation clumps or young separate
galaxies; they are whole galaxies with relatively old populations of
stars and superposed star formation regions. A similar conclusion was
made for one case before, the clump cluster UDF 6462, whose rotation
curve is dominated by a smooth component suggestive of a distorted, but
single, disk (Bournaud et al. 2008).

The UDF reveals that a high fraction of resolved, high-redshift objects
are clumpy. Among disky types, namely clump-clusters, chains, and
spirals, the comoving density of the the first two combined exceeds
that of the spirals by a factor of 2.3 beyond $z=1$ (from Fig. 6 in
Elmegreen et al. 2007a). Considering that some spirals may be too red
to see in the ACS, we can only say that the clumpy types appear to
dominate the spirals by up to a factor of $\sim2$ in number density.
Still, clump clusters and chains can be observed out to $z\sim5$ in the
ACS, which is the expected limit from bandshifting. Most likely they go
further because their $V/V_{max}$ distribution is relatively flat out
to this limit (meaning that their redshift edge has not been seen --
Elmegreen 2007a). It follows that clump clusters and chains appear to
represent an early phase of disk formation that a high fraction of
spirals go through. Our results in Section \ref{sect:results} confirm
this: The ratio of individual clump mass to galaxy mass is $\sim2$\%
for clump clusters and chains. Because a typical galaxy of this type
has $\sim5$ big clumps, this means that the ratio of total clump mass
to galaxy mass is $\sim10$\%. The ratio could have been larger in the
past as the galaxy built up. If the underlying red disk is made from
multiple episodes of clump formation, then there should be 5 to 10 of
these episodes to make the disks we observe. Also, the ratio of clump
age to bulge age is $\sim0.16$ for spiral galaxies (from Fig.
\ref{fig:nicmos_bulgeclump_histograms} and Table 1).  The age ratio is
$\sim0.6$ for clump clusters and chains. Thus there is approximately
enough time for the 5 to 10 episodes determined by the mass fraction to
convert a clump cluster or chain galaxy into a spiral galaxy.

The relative masses and ages for clumps and bulges or bulge-like clumps
support this evolutionary model.  Clumps and BLCs are more similar to
each other in clump cluster and chain galaxies than the clumps are to
bulges in spiral galaxies. In spirals, the bulges are significantly
older than the clumps ($\times6$), and significantly more massive
($\times20$ to 30; Tab. 1), whereas in the clump clusters and chains,
the BLCs are only slightly older than the clumps ($\times1.7$) and only
$\sim5\times$ more massive (Fig.
\ref{fig:nicmos_bulgeclump_histograms}, Tab. 1). Moreover, only
$\sim30$\% of the chains and $\sim50$\% of the clump clusters have
obvious BLCs on infrared images.  All of this suggests that clump
clusters and chains represent a young phase of spiral galaxy formation,
a phase that includes both bulge and disk formation. The observations
also suggest that the BLCs in clump cluster and chain galaxies are
related to the star formation regions because the two types of clumps
are similar. This supports a model where the disk forms by the
dissolution of clumps (Elmegreen et al. 2005b), and some bulges form by
the mergers of clumps (e.g., Noguchi 1999; see recent discussions in
Bournaud et al. 2007, Elmegreen et al. 2008b).

The star formation process in high redshift disks appears to be similar
to that in local spiral galaxies, i.e., an initiation by gravitational
instabilities in disk gas, followed by a cascade inside each giant
condensation to star clusters and associations. The giant clumps we see
at high redshift are presumably the first (largest) step in this
process. They probably contain unresolved subclusters. In local
spirals, the large-scale gas clumps in the spiral arms of
density-waves, or the flocculent patches in galaxies without density
waves, are the first step. These local clumps contain $10^7\;M_\odot$
in gas (Elmegreen \& Elmegreen 1983) and $10^5\;M_\odot$ in stars in
the form of clusters and OB associations, which combine to make star
complexes (Efremov 1995). At high redshift, the gas mass in each clump
is unobserved but the stellar mass is much larger than locally,
$10^7\;M_\odot$ or more (Fig.
\ref{fig:nicmos_bulgeandclumps_mass_vs_z_alltypes}). This suggests that
the gas mass in clumps at high redshift is around $10^9\;M_\odot$. If
these giant clumps result from gravitational instabilities, then the
unstable Jeans mass is this order of magnitude as well. The sizes of
star-forming regions also scale up for high redshift galaxies, from
$\sim600$ pc for complexes locally (Efremov 1995) to $\sim1.8$ kpc in
clump clusters and chains (Elmegreen \& Elmegreen 2005).

The implications of this scale-up at high redshift can be assessed from
the characteristic wavelength and mass of the gravitational
instability. The wavenumber for fastest growth in an infinitely thin
layer of column density $\Sigma$ and velocity dispersion $a$ is $k=\pi
G\Sigma/a^2$. Using half the corresponding wavelength,
$\lambda/2=\pi/k=a^2/G\Sigma$ as a measure of clump radius, and
$\Sigma\pi\left(\lambda/2\right)^2=\pi a^4/G^2\Sigma$ as a measure of
clump mass, we see that the scale up in mass by a factor $\sim100$ and
the scale up in size by a factor of $\sim3$ correspond to an increase
in $a^4/\Sigma$ by $\times100$ and an increase in $a^2/\Sigma$ by
$\times3$. Thus $a^2$ alone has to increase by $100/3\sim30$, which
means the velocity dispersion has to increase by $\times5.5$, making it
$30$ or $40$ km s$^{-1}$ instead of the local $6$ or 7 km s$^{-1}$.
This requirement is satisfied by the observation of high velocity
dispersions in high redshift galaxies (F{\"o}rster Schreiber, et al.
2006; Weiner et al. 2006), particularly in the clump cluster studied by
Genzel et al. (2006).  The scale-up also implies that $\Sigma$ has to
be larger by $100/3^2\sim10$ at high redshift. Locally, the gaseous
disks of spiral galaxies have column densities of $\sim10\;M_\odot$
pc$^{-2}$, which makes the Jeans mass $2\times10^7\;M_\odot$,
comparable to the local gas clump mass for $a=6$ km s$^{-1}$. At high
redshift, the gas column density in the disk has to be larger, perhaps
$100\;M_\odot$ pc$^{-2}$, which is comparable to the total mass column
densities of the inner regions of today's spirals.  This equivalence in
column density is sensible considering that the UDF galaxies in our
survey are smaller than local galaxies by about a factor of $\sim2$
(Elmegreen et al. 2005a). Thus, we are probably witnessing the buildup
of the inner disks and bulges of today's spiral galaxies by
star-formation processes very similar to today's but with relatively
large turbulent speeds.

The star formation rate in the clumps can be estimated from the masses
and ages. Considering the average values in Figures
\ref{fig:nicmos_bulgeandclumps_mass_vs_z_alltypes} and
\ref{fig:nicmos_bulgeandclumps_age_vs_z_alltypes}, namely,
$10^{7.5}\;M_\odot$ for clump mass and a broad range around $\sim10^2$
Myr for clump age, we get an average rate of $\sim0.3\;M_\odot$
yr$^{-1}$ for each clump. If we set this equal to $\epsilon M_{gas}
\left(G\rho_{gas}\right)^{1/2}$ for efficiency in a free-fall time,
$\epsilon$, clump gas mass, $M_{gas}=10^{9}M_\odot$, and clump core
density $\rho_{gas}$, which is $M_{gas}/L_{clump}^3$ for 3D clump size
$L_{clump}\sim1$ kpc, we get an efficiency of $\epsilon\sim$ few
percent. This means that a typical initial gas clump of
$10^{9}\;M_\odot$ converts a few percent of its gas into stars on each
dynamical time, which is $\sim20$ Myr. Such an efficiency is comparable
to that for local star complexes.

\section{Conclusions}
\label{sect:conc}

We have measured the magnitudes and colors of star-forming clumps,
bulge-like clumps (BLCs) and bulges in all of the suitable clump
clusters, chain galaxies, and spiral galaxies larger than 0.3 arcsec in
the UDF. Measurements were made on the ACS and NICMOS images and the
colors were determined at consistent angular resolutions. The
observations were fitted to population synthesis models to obtain clump
masses, ages, star formation decay times in an exponentially declining
rate, and extinctions. Approximately 30\% of the chain galaxies and
50\% of the clump clusters contain bright red clumps that could be a
young bulge. These bulge-like clumps are not always centralized:
$\sim20$\% are significantly off-center, sometimes in a ring.

The results suggest that clump clusters and chain galaxies are young
spiral galaxies.  The BLCs are relatively young and low-mass in clump
clusters and chain galaxies: older than the clumps by a factor of only
$\sim1.7$, and more massive by a factor of $\sim5$. In contrast, the
bulges in spirals are $\sim6$ times older than the clumps and
$20-30\times$ more massive. All of the systems appear to contain older
stars in an interclump component. From the relative masses of the
clumps in clump clusters and chains ($\sim10$\%), and from their
relative ages ($\sim15$\%) we infer that the clump-cluster and chain
galaxy phase lasts for $\sim5-10$ clump formation epochs, which should
be several Gyr considering the clump ages.

High redshift disks appear to form stars by gravitational instabilities
in their gaseous components. The resulting star-formation clumps are
$3\times$ larger and $\sim100\times$ more massive than today's star
complexes, but both seem defined by the disk Jeans length and mass. The
scale-up at high redshift could result from a factor of $\sim5$
increase in turbulent speed and a factor of $\sim10$ increase in gas
mass column density. Such a turbulent speed is consistent with
observations, and the high gas column density is consistent with the
total column density of today's inner disks. The initiation of star
formation is probably related to the origin of the gas, which is
unobserved. The gas could arrive by accretion in a relatively smooth
three-dimensional flow, or it could move in two-dimensions from the
outer disk following an interaction. In either case, we appear to be
witnessing the formation of the bulges and inner disks of modern
galaxies.

We gratefully acknowledge Vassar College for student support for M.F.
and J.L. and for publication support through a Research Grant. We are
grateful to Jennifer Mack for providing ACS filter and HST throughput
functions.  Helpful comments by the referee are appreciated. This
research has made use of the NASA/IPAC Extragalactic Database (NED)
which is operated by the Jet Propulsion Laboratory, California
Institute of Technology, under contract with the National Aeronautics
and Space Administration.

\clearpage
\begin{table}
\footnotesize
\begin{center}
\caption{Mean $\pm$ Standard Deviations for the Distributions Comparing
Clumps (C) and Bulges (B) with Various Metallicities}
\end{center}
\begin{tabular}{ccccc}
\tableline\tableline

Comparison &  Metallicity & Spirals  & Chains & Clump Clusters  \\
$\log\left(M_{clump}/M_{bulge}\right)$ & 0.008 & $-1.42\pm1.05$ & $-0.38\pm0.49$ & $-0.73\pm0.60$\\
                                       & 0.004 & $-1.46\pm1.08$ & $-0.33\pm0.50$ & $-0.77\pm0.62$\\
                         & 0.008 (B) 0.004 (C) & $-1.48\pm1.06$ & $-0.43\pm0.48$ & $-0.73\pm0.60$\\
                         & 0.004 (B) 0.008 (C) & $-1.41\pm1.07$ & $-0.28\pm0.51$ & $-0.76\pm0.62$\\
\tableline
$\log\left(Age_{clump}/Age_{bulge}\right)$ & 0.008 & $-0.80\pm0.87$ &  $0.19\pm0.75$ & $-0.24\pm1.07$\\
                                           & 0.004 & $-0.81\pm0.90$ & $-0.39\pm0.64$ & $-0.25\pm1.10$\\
                             & 0.008 (B) 0.004 (C) & $-0.84\pm0.88$ & $-0.0093\pm0.59$ & $-0.32\pm1.10$\\
                             & 0.004 (B) 0.008 (C) & $-0.77\pm0.90$ & $-0.18\pm0.59$ & $-0.16\pm1.10$\\
\tableline
$\log\left(\tau_{clump}/\tau_{bulge}\right)$ & 0.008 & $0.18\pm0.66$ & $0.05\pm0.39$ & $0.001\pm0.50$\\
                                             & 0.004 & $0.18\pm0.63$ & $0.02\pm0.48$ & $-0.06\pm0.49$\\
                               & 0.008 (B) 0.004 (C) & $0.15\pm0.66$ & $-0.002\pm0.40$ & $-0.007\pm0.48$\\
                               & 0.004 (B) 0.008 (C) & $0.22\pm0.63$ & $0.06\pm0.47$ & $-0.06\pm0.51$\\
\tableline
$\log\left(Ext_{clump}/Ext_{bulge}\right)$ & 0.008 & $-0.46\pm2.05$ & $-1.38\pm1.79$ & $-0.85\pm3.07$\\
                                             & 0.004 & $-0.76\pm1.96$ & $-0.24\pm2.45$ & $-1.11\pm2.90$\\
                               & 0.008 (B) 0.004 (C) & $-0.56\pm1.95$ & $-1.26\pm1.88$ & $-0.86\pm3.02$\\
                               & 0.004 (B) 0.008 (C) & $-0.66\pm2.03$ & $-0.36\pm2.17$ & $-1.10\pm2.91$\\
\tableline
\end{tabular}
\end{table}


\clearpage


\clearpage


\clearpage


\clearpage


\clearpage


\clearpage
\begin{figure}\epsscale{1}
\plotone{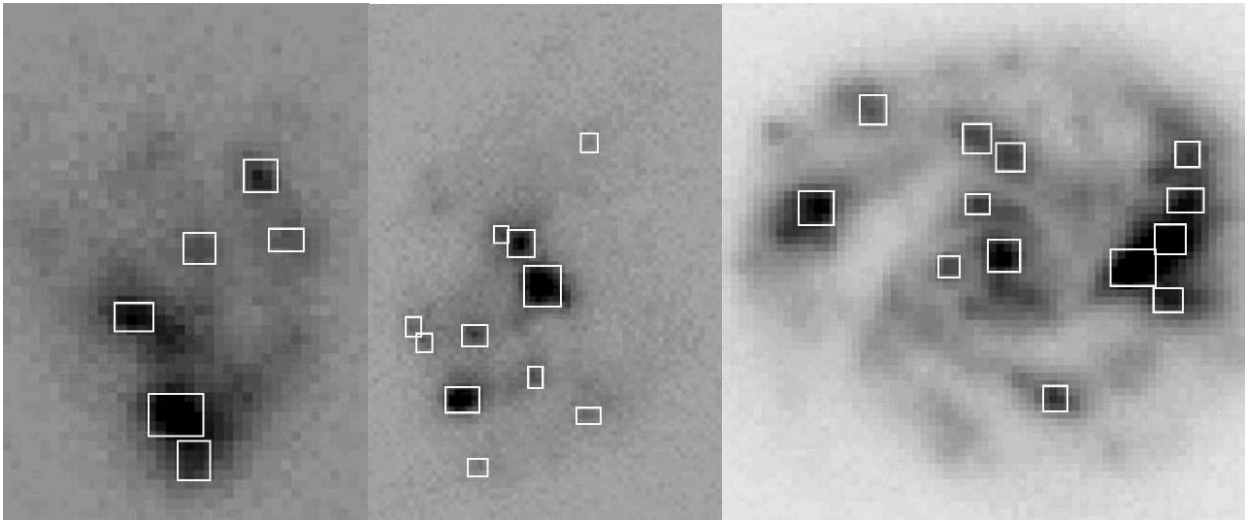} \caption{Boxes used for magnitude determinations of
the clumps in three galaxies, UDF 5827, UDF 1666, and UDF 3372, which
were also shown in previous figures (Figs. 5, 1, and 1, resp.). The
images are $i_{775}$ band.
}\label{fig:UDF2510,1666,3372linear}\end{figure}

\clearpage
\begin{figure}\epsscale{.9}
\plotone{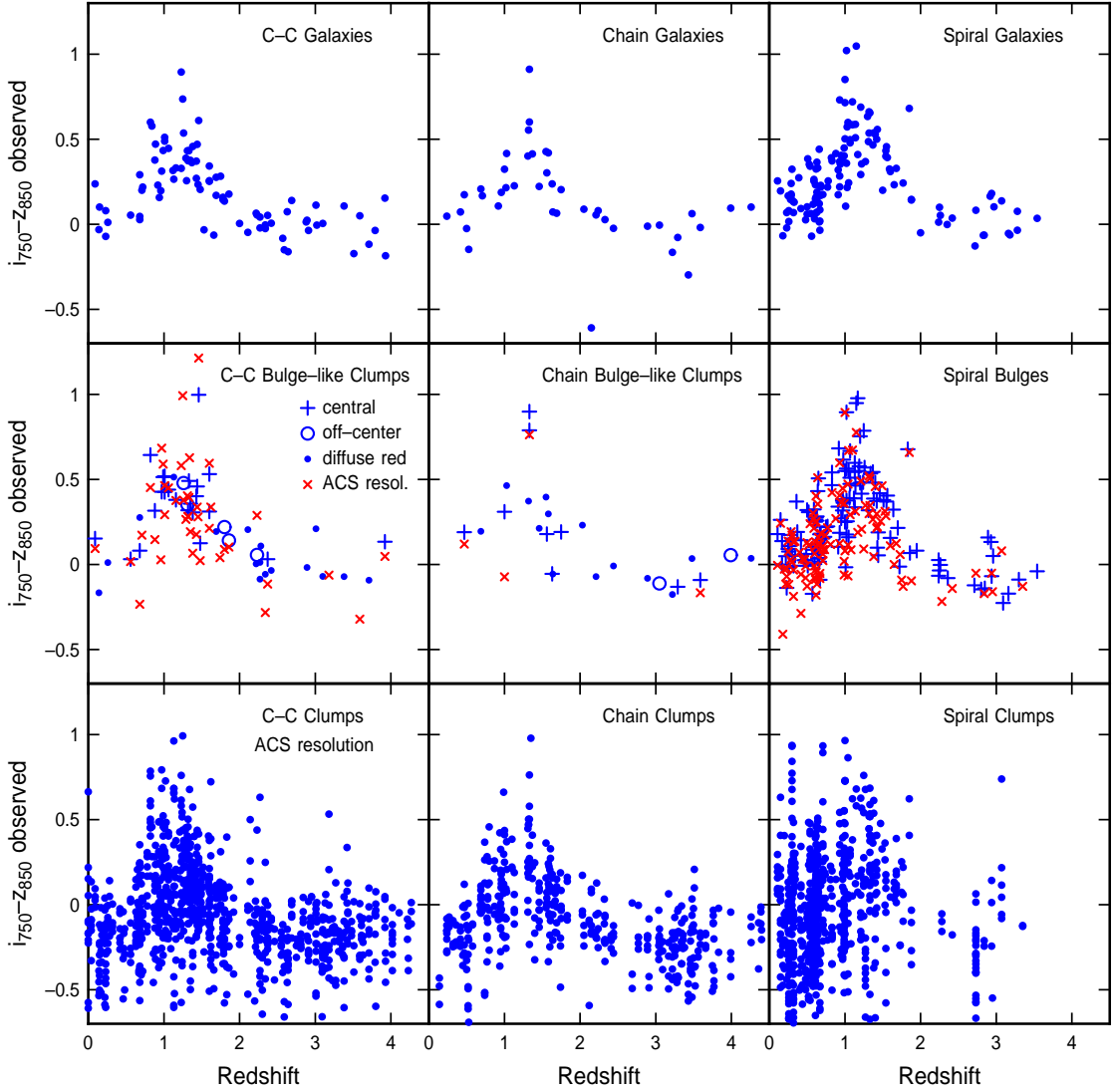} \caption{Observed colors $i_{750}-z_{850}$ versus
redshift. The top row of panels plots colors of whole galaxies from the
NICMOS online catalog. The second row plots in blue the colors of
centralized (plus signs) and off-center (circles) bulges/BLCs from the
ACS images that were convolved to the NICMOS resolution, which is three
times larger than that of the ACS camera.  The blue dots in the middle
row are colors for inner bright regions that we do not consider to be
bulges, measured off the ACS with NICMOS resolution. The red crosses
show the bulges/BLCs measured whenever possible with the full
resolution ACS images. The bottom row shows clump measurements,
including both bulge, BLC, and non-bulge clumps, all with ACS
resolution. The peak at $z=1.3$ is the restframe Balmer jump.}
\label{fig:nicmos_I-ZvsZwithclumps}
\end{figure}

\clearpage
\begin{figure}\epsscale{.9}
\plotone{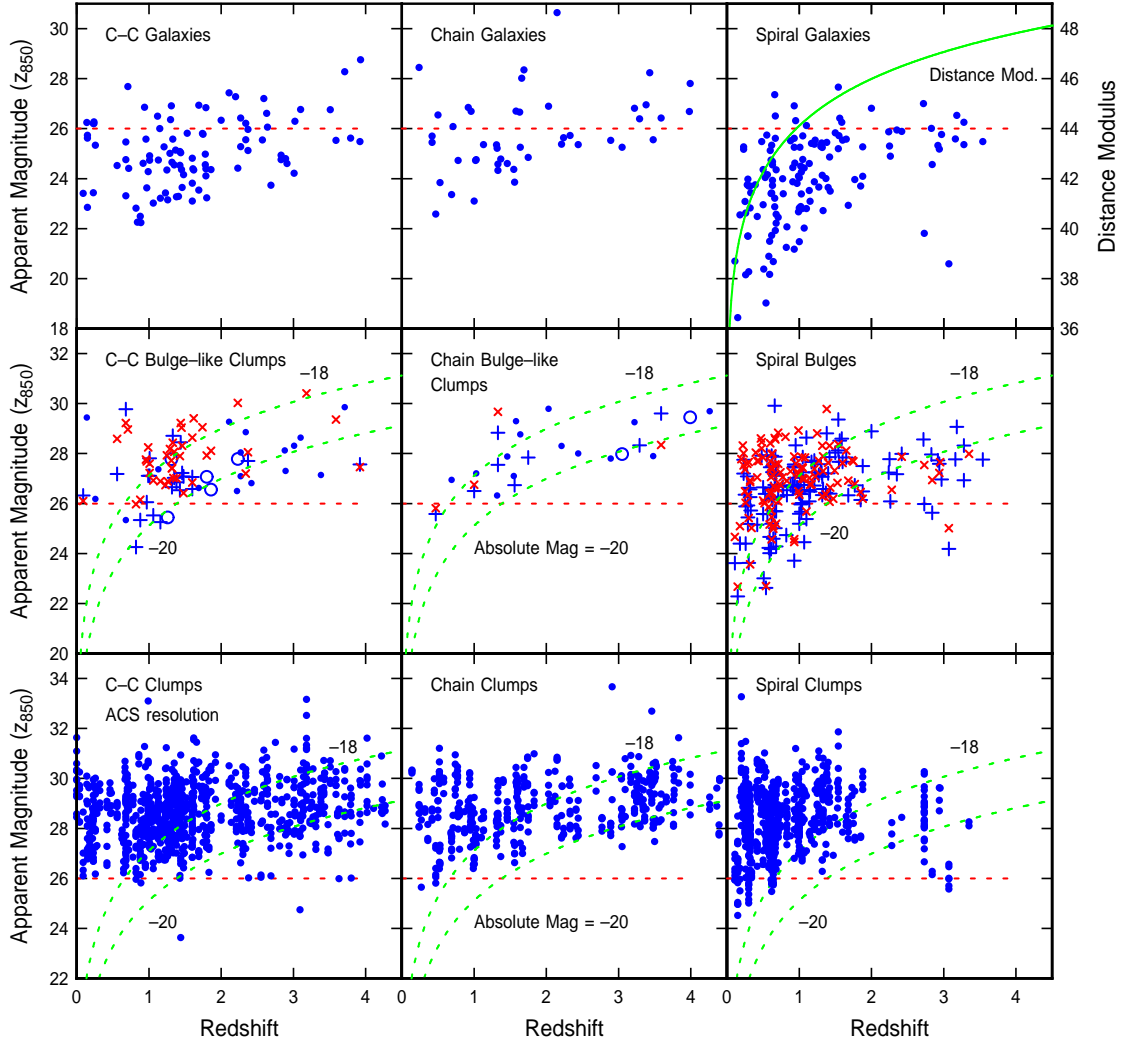} \caption{Apparent $z_{850}$ band magnitudes for the
same galaxies, bulges, BLCs, and clumps as in Fig. 6, with the same
symbol convention. The horizontal red dotted lines indicate a magnitude
of 26 for all panels, to aid comparisons. Galaxies are typically 2 to 3
magnitudes brighter than the bulges, which are 1 to 2 magnitudes
brighter than the clumps. The green curve in the upper right panel is
the distance modulus as a function of redshift. The dotted green curves
in the middle and bottom rows indicate absolute magnitudes of $-18$ and
$-20$. The bulge/BLC magnitudes increase with redshift and the clump
magnitudes are about constant. Considering that the bulge/BLC and clump
masses are more constant than this (Fig. 14), the trends here are
probably from bandshifting.} \label{fig:nicmos_appmag_vs_z}\end{figure}

\clearpage
\begin{figure}\epsscale{1}
\plotone{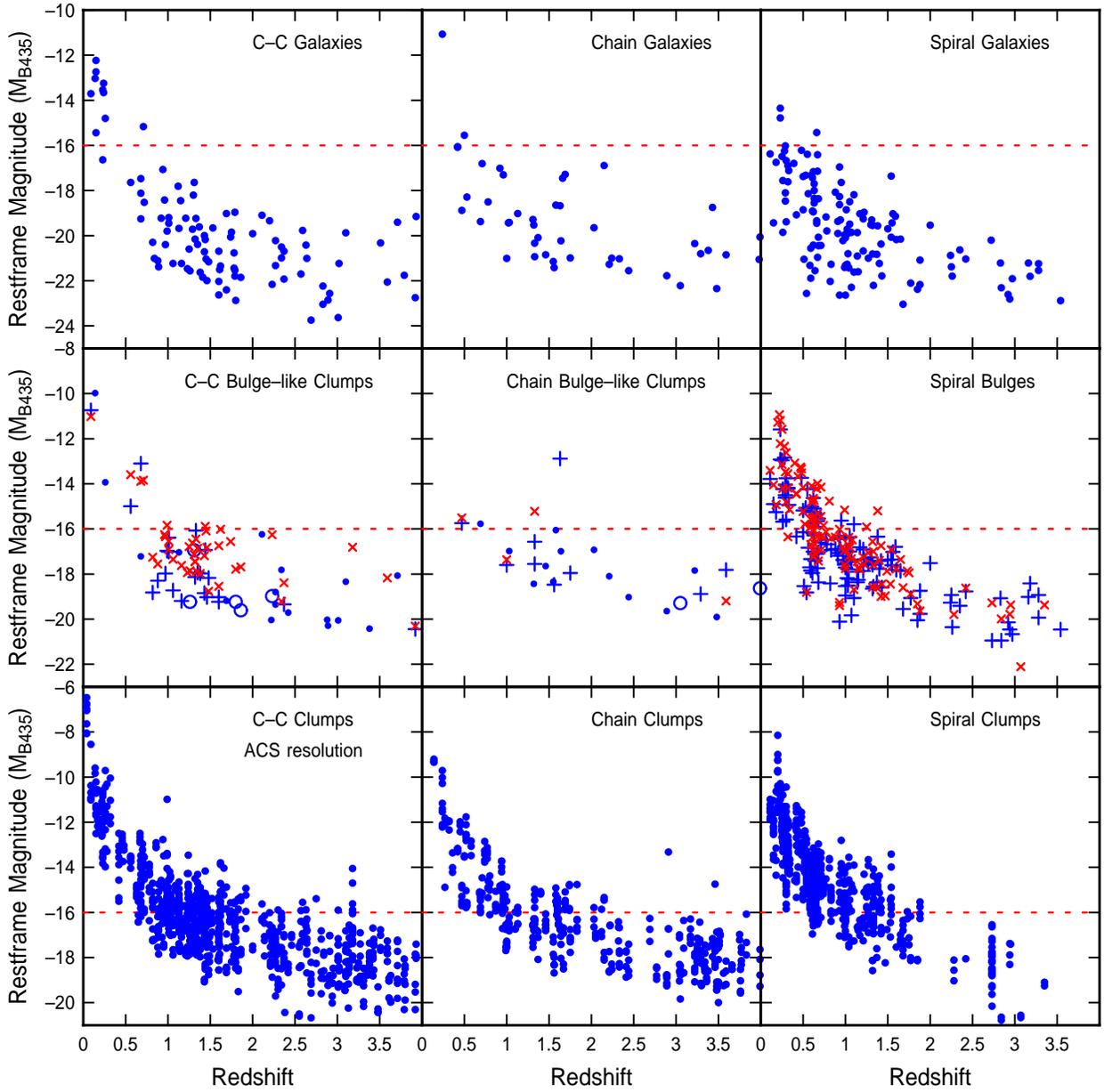} \caption{Absolute magnitudes in the restframe
$B_{435}$ band (and observed $z_{850}$ band at $z>0.95$) for whole
galaxies, bulges and clumps. Symbol types are the same as in the
previous two figures. The red dashed line indicates an absolute
magnitude of $-16$ for each panel, to aid in comparisons.  These
restframe magnitudes were determined by interpolation between ACS
bands. Beyond a redshift of 0.95, the restframe $B_{435}$ shifts past
the center of the $z_{850}$ band and interpolation is not possible for
ACS resolution; all points at higher redshifts take the absolute
$z_{850}$ magnitude as a best estimate.}
\label{fig:nicmos_absmag_vs_z}\end{figure}

\clearpage
\begin{figure}\epsscale{1}
\plotone{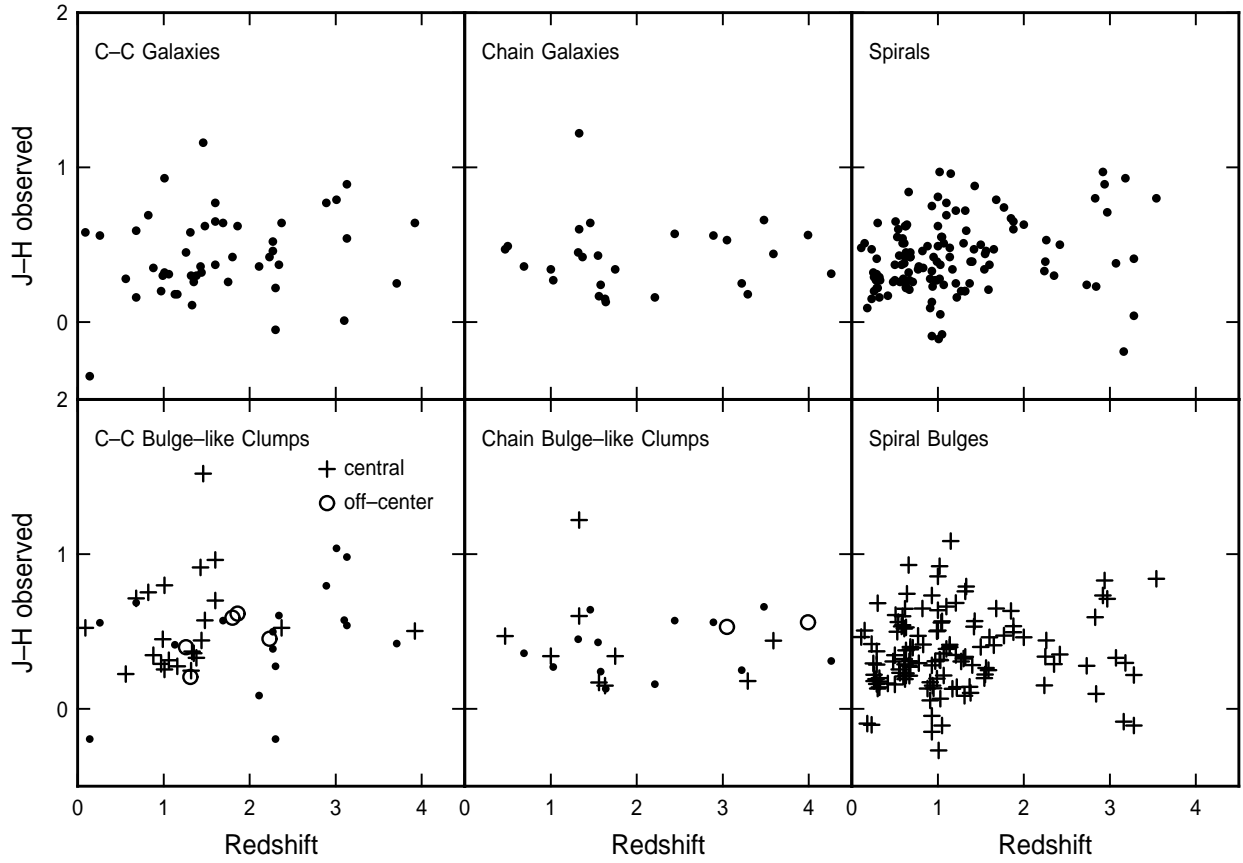} \caption{NICMOS J-H colors versus redshift for whole
galaxies and bulges/BLCs. }\label{fig:nicmos_J-HvsZ}\end{figure}

\clearpage
\begin{figure}\epsscale{1}
\plotone{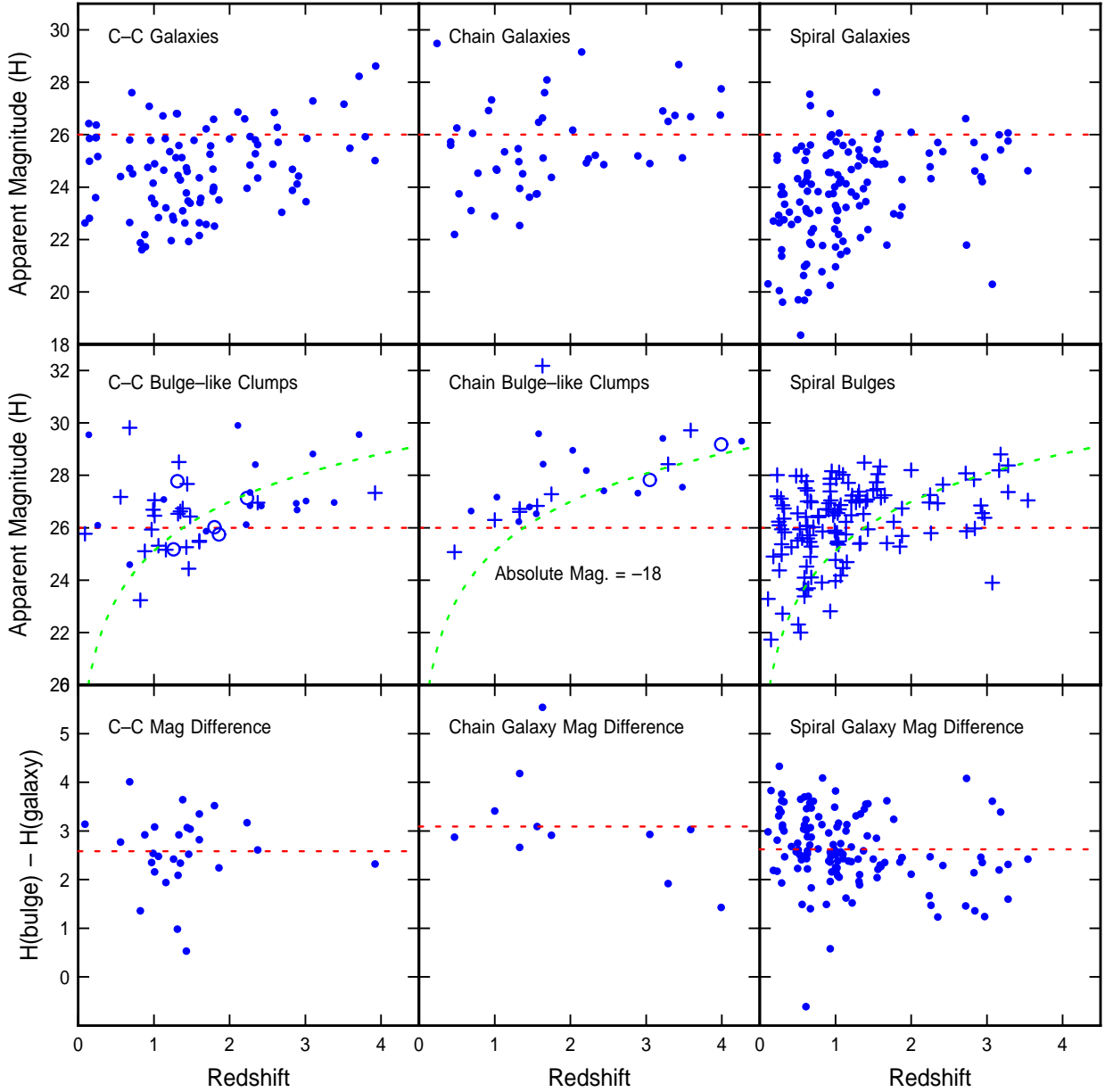} \caption{Apparent H-band magnitudes for whole
galaxies and bulges/BLCs, with the differences between these two
measurement in the bottom panels. The horizontal dotted lines in the
top two rows are fiducial markers of magnitude 26, and the dotted lines
in the bottom row are averages for the plotted clump magnitudes. If the
magnitude difference in H band is approximately 2.5 times the log of
the mass ratio, then the bulge-to-galaxy mass ratios are 0.09, 0.06,
and 0.09 for clump clusters, chains and bulges, respectively, with an
uncertainty of a factor of $\sim2$. The rising dotted curves in the
middle panels correspond to an absolute magnitude of $-18$.
}\label{fig:nicmos_appmagH_vs_z}\end{figure}

\clearpage
\begin{figure}\epsscale{1}
\plotone{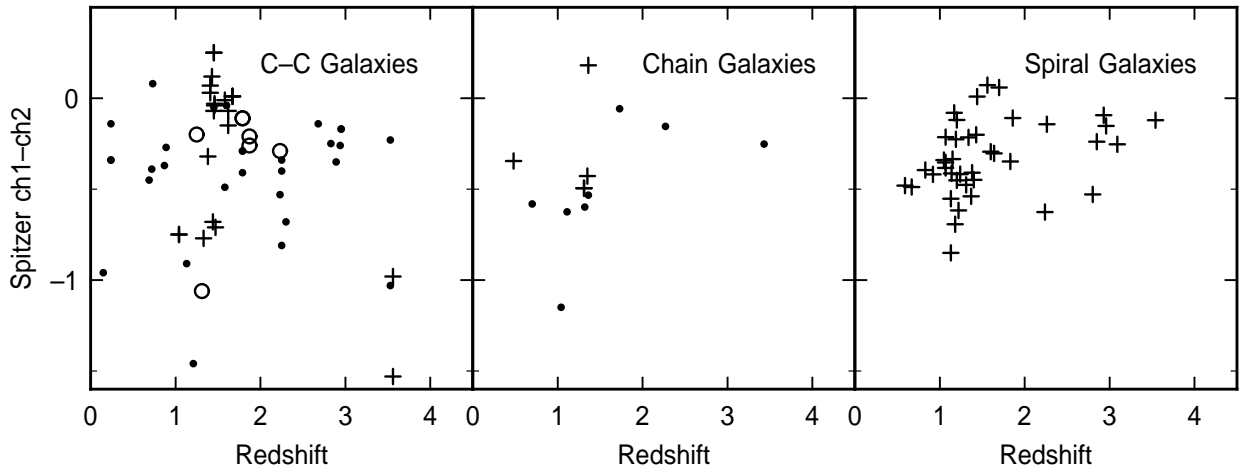} \caption{Colors for whole galaxies from Spitzer Space
Telescope channel 1 magnitude (3.6$\mu m$) minus channel 2 magnitude
(4.5$\mu m$), versus the redshift. For clump clusters, the galaxies
with centralized bulges (plus signs) could be slightly redder than the
galaxies with offset bulges (circles), but there is a lot of scatter.
Most bulge/BLC galaxies are redder than non-bulge galaxies (dots). The
Spitzer Space Telescope does not resolve the bulges or
clumps.}\label{fig:nicmos_spitzer}\end{figure}

\clearpage
\begin{figure}\epsscale{1}
\plotone{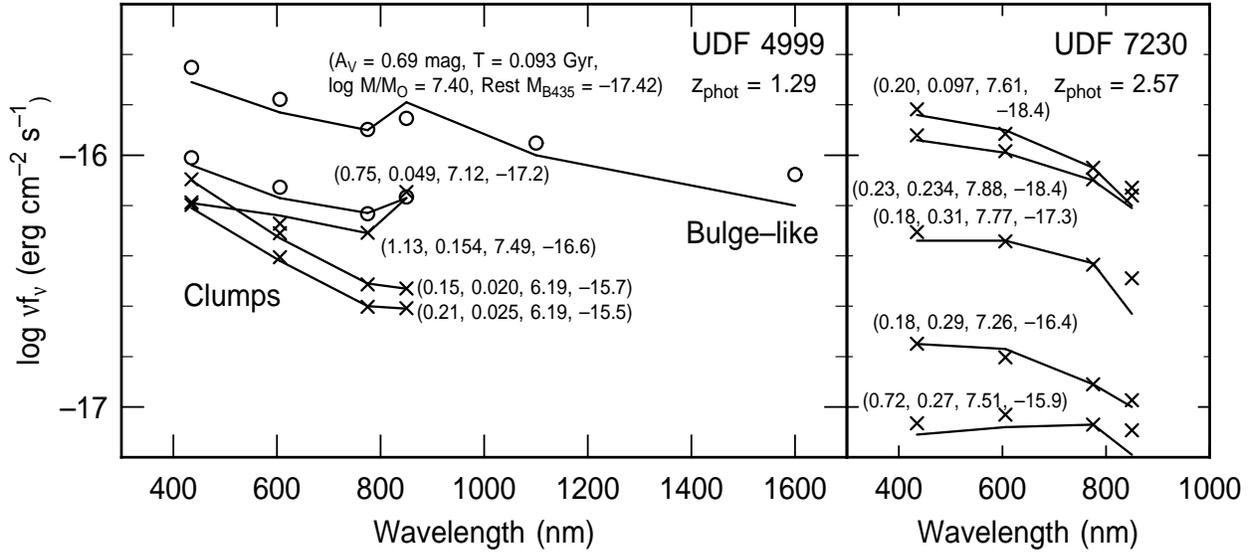} \caption{SED fits to two galaxies depicted in Figs. 1
and 2.  The NICMOS BLC in UDF 4999 is shown here as a long line with
points at 1100 and 1600 nm. The visible band magnitudes for it were
measured from the ACS image deconvolved to the NICMOS resolution. UDF
7230 has no obvious NICMOS bulge or BLC. The parameters near each fit
indicate the preferred extinction in restframe visible light, the age
of the region, the mass, and the restframe absolute $B_{435}$
magnitude. For UDF 4999, the uppermost short line with circle markers
is the bulge measured in the high-resolution ACS image. All other
clumps were measured with ACS resolution too. The ACS magnitudes are
fainter than the convolved magnitudes in the long line because the high
resolution of the ACS can better isolate the bulge/BLC.
}\label{fig:nicmos_colors_2601_3209}\end{figure}

\begin{figure}\epsscale{1}
\plotone{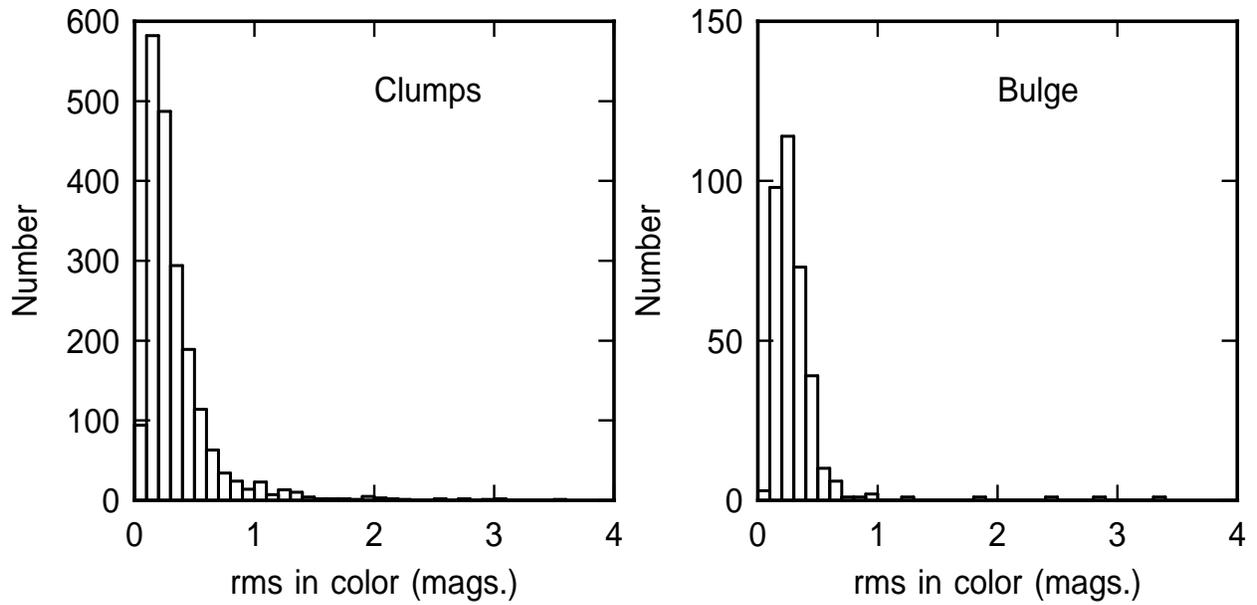} \caption{Histogram of the rms deviations between the
model colors and the observed colors for clumps (left) and bulges
(right).  The color deviation is calculated as the square root of the
sum of the squares of the differences between the model and observed
colors for all available colors.}\label{fig:nicmos_plotrms}\end{figure}

\clearpage
\begin{figure}\epsscale{.8} \plotone{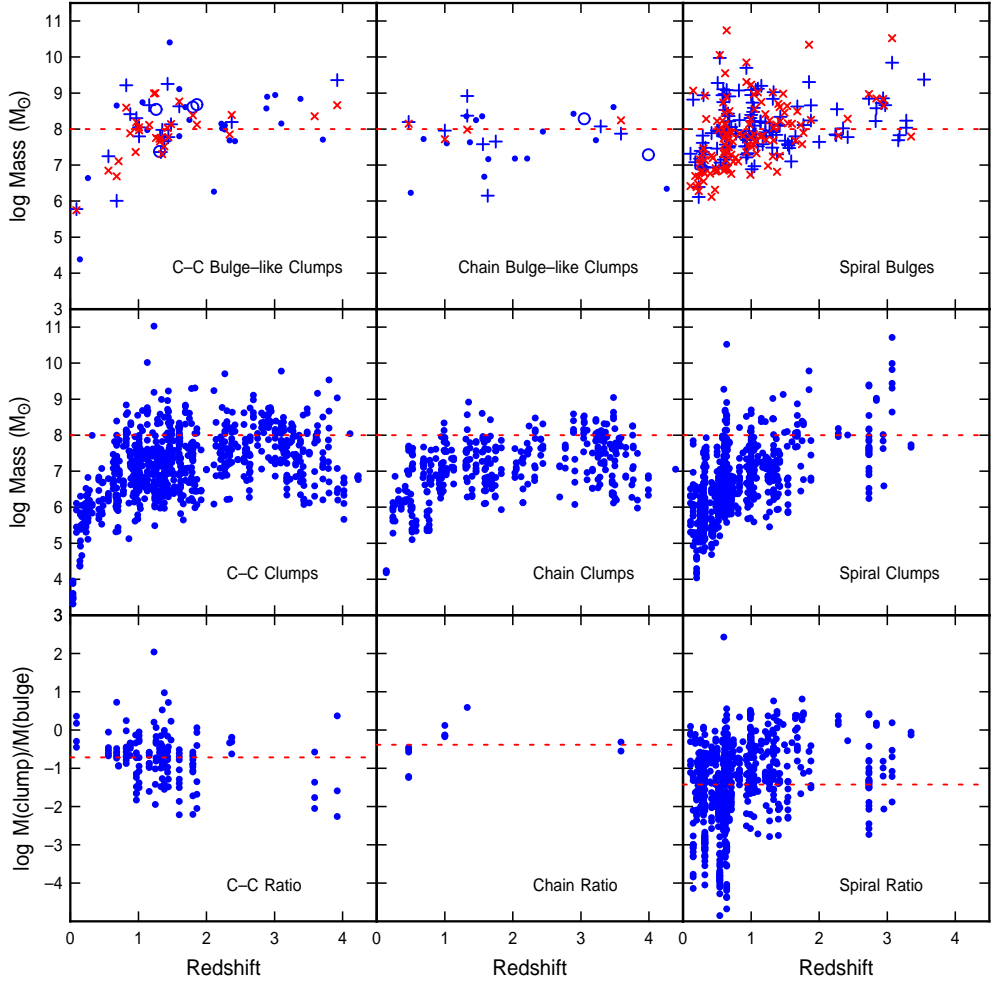} \caption{The bulge, BLC, and
clump masses obtained from the SED fits are shown as functions of the
redshift in the top two rows, and the log of the ratio of these masses
is shown in the bottom row. In the top row, plus signs are central
bulges or BLCs, circles are offset BLCs, and dots are inner diffuse
regions not resembling a bulge, all measured with NICMOS resolution,
even in the ACS bands, while the red crosses are bulges/BLCs measured
only in the optical bands on the high resolution ACS image and fitted
to the models only with those bands.  The red dotted lines in the top
two rows indicate a mass of $10^8\;M_\odot$ for comparisons. The dotted
red line in the bottom row indicates the average value for each galaxy
type. The bulge/BLC and clump masses are about constant with redshift
beyond $z\sim0.5$. A typical bulge/BLC has a mass between $10^8$ and
$10^9\;M_\odot$, slightly lower in chains that in clump clusters and
spirals. A typical clump has a mass of $10^7$ to $10^8\;M_\odot$ for
$z>0.5$. Chains and clump clusters have larger clumps compared to their
BLCs than spiral galaxies, suggesting a more primitive state for the
former types. High redshift spirals also have larger clumps compared to
their bulges than low redshift spirals. Uncertainties for each point
are plotted in Fig. 18.}
\label{fig:nicmos_bulgeandclumps_mass_vs_z_alltypes}\end{figure}

\clearpage
\begin{figure}\epsscale{1}
\plotone{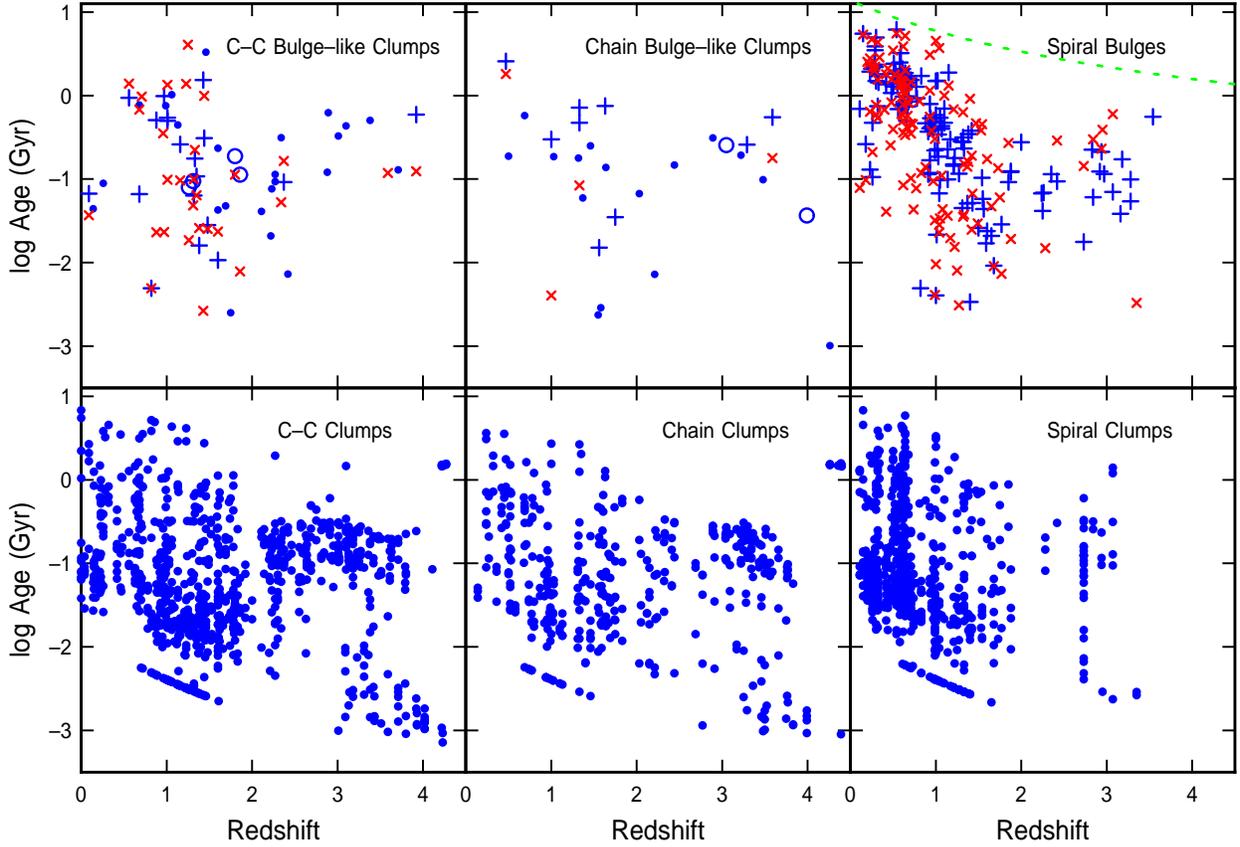} \caption{The bulge, BLC, and clump ages obtained from
SED fits are shown versus the redshift. Symbol types are as in previous
figures. The dotted green line in the top right is the age of the
universe observed from the restframe of the galaxy. Age determinations
are more uncertain than mass determinations because of a greater
sensitivity of the age to the extinction. Uncertainties for each point
are plotted in Fig. 18. The trend observed here is sensible. Bulge/BLC
ages are typically several Gyr less than the age of the universe, and
some appear to be young in their restframe. Clumps are significantly
younger than bulges only for spiral galaxies (see also Fig. 18).
}\label{fig:nicmos_bulgeandclumps_age_vs_z_alltypes}\end{figure}

\clearpage
\begin{figure}\epsscale{1}
\plotone{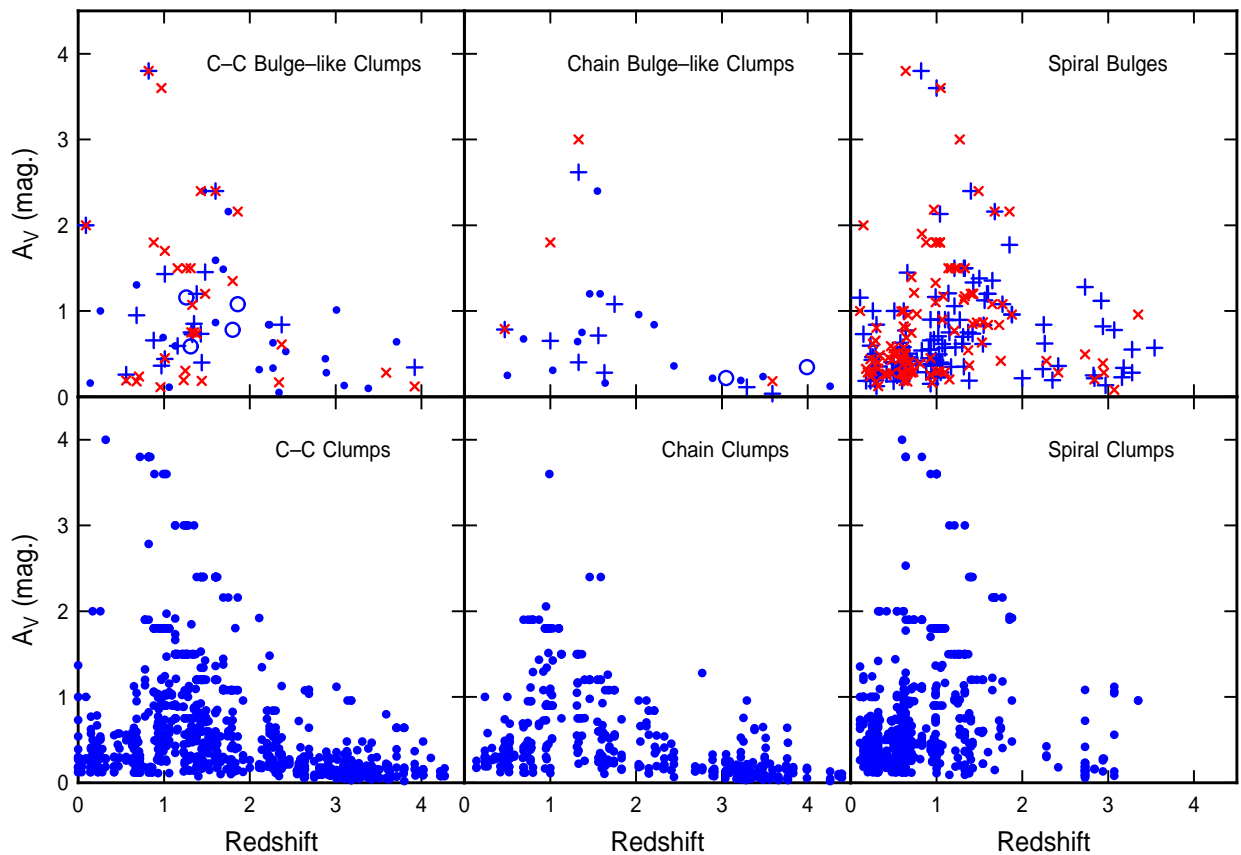} \caption{The bulge, BLC, and clump extinctions
obtained from SED fits are shown versus the redshift. Symbol types are
as in previous figures. The extinction decreases systematically with
redshift, reflecting a possible change in metallicity. This decrease
reinforces the age decrease seen in the previous figure because the
age-extinction degeneracy has the opposite trend. No extinction
differences are observed among the various clump and galaxy types.
}\label{fig:nicmos_bulgeandclumps_ext_vs_z_alltypes}\end{figure}

\clearpage
\begin{figure}\epsscale{.8}
\plotone{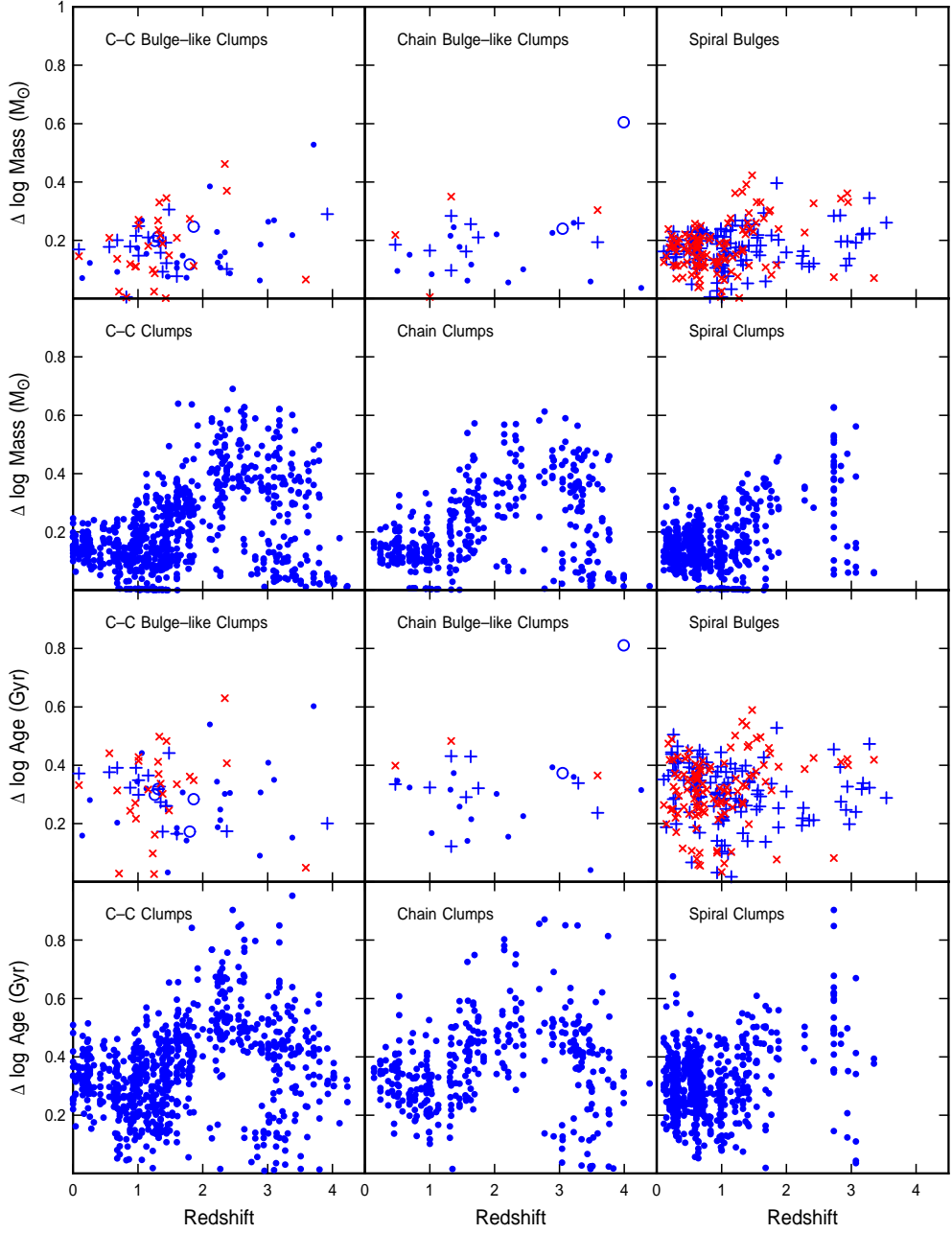} \caption{Root mean squared deviations of the bulge,
BLC, and clump masses (top) and ages (bottom) obtained from SED fits
are shown versus the redshift. Symbol types are as in the previous
figures.  These rms values are the uncertainty in the previously
plotted points which were too congested in the figures to show error
bars. The typical uncertainty for bulge mass is a factor of $\sim1.4$
and for clump mass is a factor of 1.6 to 2. The typical uncertainty for
bulge age is a factor of $\sim2$ and for clump age is a factor of
$\sim2.5$. These uncertainties reflect the range of fitted values for
these quantities among all of the fits with the closest matches to the
observed colors.
}\label{fig:nicmos_bulgeandclumps_age_vs_z_alltypes_rms}\end{figure}

\clearpage
\begin{figure}\epsscale{1}
\plotone{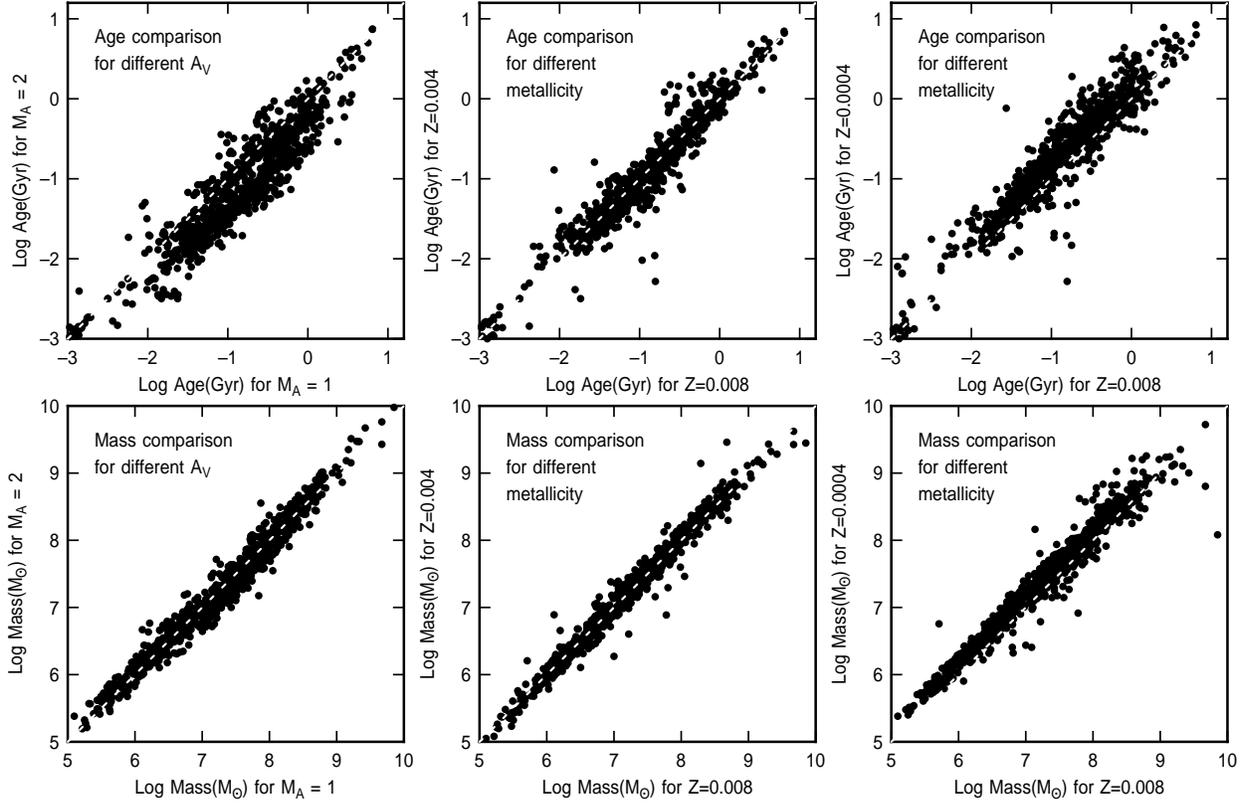} \caption{Three comparisons of fitted age and mass for
clumps in clump cluster galaxies. On the left, the fits for extinctions
that differ by a factor of two are compared. In the middle and right,
the fits for metallicities that differ by a factor of 2 and a factor of
20, respectively, are compared. Extinction values that are too high
produce ages that are too low, but they do not change the fitted mass
significantly. Metallicity values that are too low by a factor of 2
hardly change either age or mass, but metallicities that are low by a
factor of 20 increase the age and mass by $\sim50$\%. In all panels the
white dashed line indicates equality of the two fits.
}\label{fig:nicmos_clump_massM42_vs_massM52_andage2}\end{figure}

\clearpage
\begin{figure}\epsscale{1}
\plotone{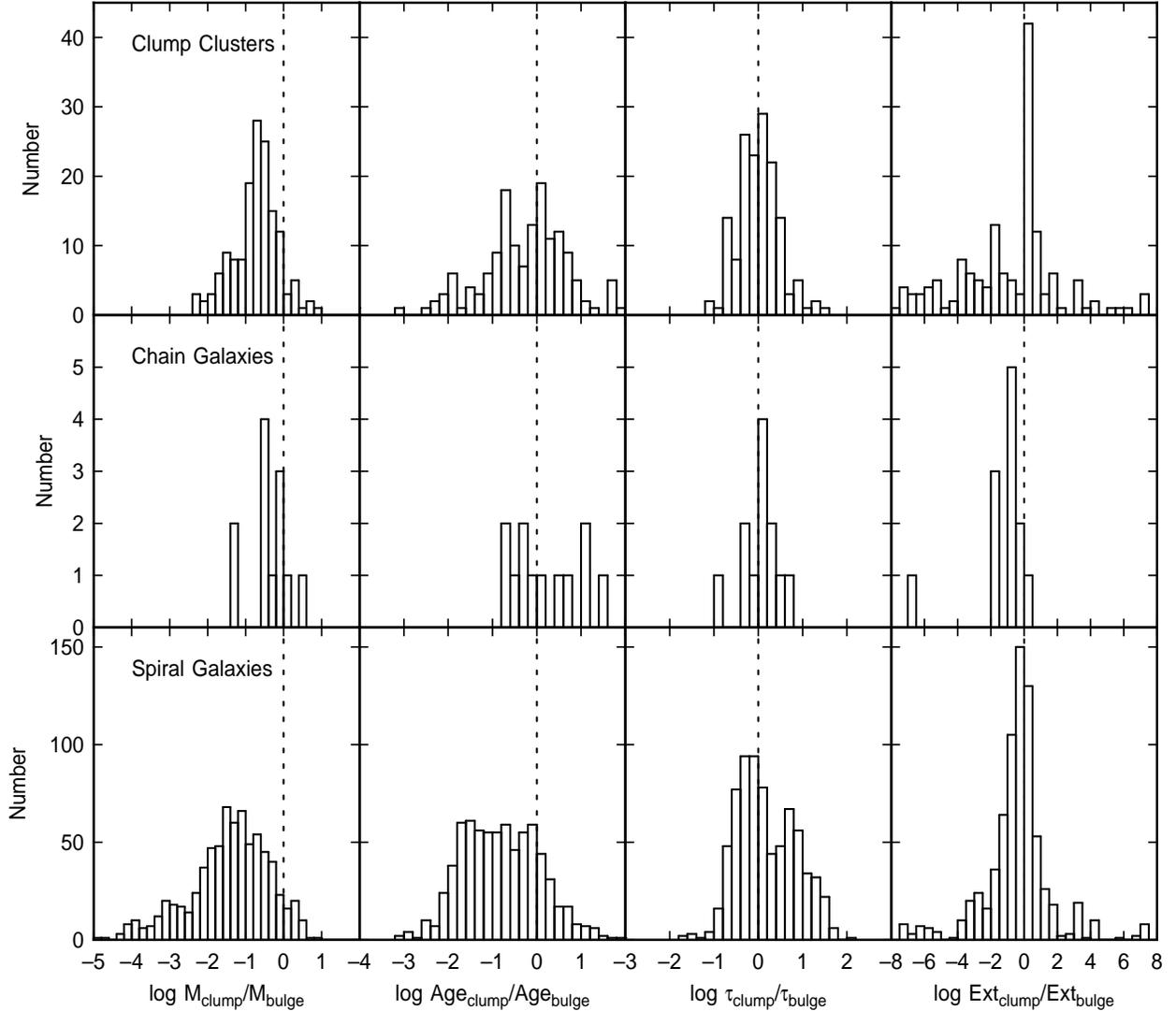} \caption{Histograms of the ratios of clump and
bulge/BLC properties obtained from the SED fits. The most significant
trends are for mass and age. The decay times for the assumed
exponential star formation rates ($\tau$) and the extinctions (``Ext'')
are about the same for clumps and bulges/BLCs in each galaxy type. The
clump masses are generally smaller than the bulge/BLC masses, although
this difference is less for clump clusters and chains than for spirals.
The clump ages are smaller than the bulge ages for spirals. These two
trends suggest that clumps and bulge-like clumps are more similar to
each other in clump cluster and chain galaxies than the clumps and
bulges are to each other in spiral galaxies.
}\label{fig:nicmos_bulgeclump_histograms}\end{figure}

\end{document}